\documentclass[prb,letterpaper,twocolumn,showpacs,preprintnumbers,floatfix]{revtex4} 

\usepackage{amsfonts}
\usepackage{amsmath}
\usepackage{latexsym}
\usepackage{graphicx}
\usepackage{bm}
\newcommand{\Fref}[1]{Fig.~\ref{#1}}
\newcommand{\Eref}[1]{Eq.~(\ref{#1})}
\newcommand{\Rref}[1]{Ref.~\onlinecite{#1}}
\newcommand{\etal}{\textit{et~al.}}
\newcommand{\Tc}{\ensuremath{T_c}}
\newcommand{\nn}{\nonumber}
\newcommand{\sst}{\scriptstyle}

\begin{document}

\title{Temperature dependence of the specific heat and the penetration depth of
  anisotropic-gap BCS superconductors for a factorizable pairing potential}

\author{T. M. Mishonov}
\email[E-mail: ]{mishonov@phys.uni-sofia.bg}
\author{S. I. Klenov}
\email[E-mail: ]{sklenov@cablebg.net; sklenov@hotmail.com}
\author{E. S. Penev}
\altaffiliation{Present address: Institute of
Physics, University of Basel, Klingelbergstr. 82, CH-4056 Basel,
Switzerland}
\affiliation{Department of Theoretical Physics, Faculty of
Physics, Sofia University ``St.~Kliment Ohridski'',
5 J. Bourchier Boulevard, BG-1164 Sofia, Bulgaria}

\date{\today}

\begin{abstract}
An explicit expression for the temperature dependence of the specific heat of
clean anisotropic-gap superconductors is derived within the weak-coupling BCS
approximation. The specific heat is presented as a functional of the
superconducting gap on the Fermi surface. The obtained formula interpolates
between the correct low coupling jump at \Tc\ and the low temperature behavior
for $T\ll \Tc.$ For isotropic superconductors the formula gives a relation
between the specific heat and the superconducting gap. For anisotropic
superconductors, the interpolation formula incorporates averaging of powers of
the gap anisotropy function over the Fermi surface and provides a suitable set
for fitting model Hamiltonians to experimental data. The work of the
interpolation formula is illustrated by (i) the Pokrovsky formula for the
specific heat jump, (ii) Gor'kov and Melik-Barkhudarov formulas for the
Ginzburg-Landau coefficients, (iii) the Moskalenko two-band formula for the
specific heat jump, (iv) the temperature dependence of the specific heat for
the two-band model, applicable to MgB$_2$, (v) the two-dimensional $d$-wave
model, applicable for YBa$_2$Cu$_3$O$_{7-\delta}$, and (vi) the Zhitomirsky
and Rice triplet $p$-wave model with horizontal line nodes for
Sr$_2$RuO$_4$. The temperature dependence of the penetration depth is
illustrated by fitting the general theoretical formula to the experimental
data for MgB$_2$, YBa$_2$Cu$_3$O$_{7-\delta}$, and the triplet superconductor
Sr$_2$RuO$_4$.
\end{abstract}

\pacs{74.20.De, 74.25.Bt}

\maketitle

%=======================================================================
\section{Specific heat}
%=======================================================================

Virtually all recently studied superconductors exhibit considerable
anisotropy of the superconducting gap $\Delta_p\,(T)$ over the Fermi
surface $\varepsilon_p=E_F$. Despite the strong coupling effects and
influence of disorder, which are all essential as a rule, for a
qualitative analysis it is particularly useful to start with the
weak-coupling BCS approximation for clean superconductors. In this
case, very often model factorizable pairing potentials give an
acceptable accuracy for the preliminary analysis of the experimental
data.

The aim of the present work is twofold. Firstly, we shall derive an explicit
interpolation formula for the temperature dependence of the specific heat
$C(T).$ The formula is formally exact for factorizable pairing kernels which
are consequence of the approximative separation in superconducting order
parameter derived in BCS weak-coupling approximation by
Pokrovskii.\cite{Pokrovskii:62} Our formula reproduces the specific heat jump
derived by Pokrovskii\cite{Pokrovskii:62} for arbitrary weak coupling kernels
and Gor'kov and Melik-Barkhudarov\cite{Gor'kov:63} results for the
Ginzburg-Landau (GL) coefficients of an anisotropic superconductor. That is
why we believe that the suggested formula can be useful for the analysis of
experimental data when only gap anisotropy and band structure are
known. Secondly, within the same system of notions and notation we present the
recent results by Kogan\cite{Kogan:02} for the penetration depth $\lambda(T)$,
and propose for the zero-scattering case new formulas which may be used for
experimental data processing.

We begin with the entropy of a Fermi system per unit volume divided by the
Boltzmann's constant $k_B$
\begin{equation}
\label{S}
    S(T)=-2 \, \overline{n_p\ln n_p+(1-n_p)\ln(1-n_p)},
\end{equation}
where the factor 2 takes into account the spin degeneracy and the
overline denotes integration over the $D$-dimensional momentum space
\begin{equation}\label{sr}
 \overline{f_p}=\int_{-\infty}^{\infty}\cdots\int_{-\infty}^{\infty}
 \frac{d^D p}{(2\pi\hbar)^D}f(\mathbf{p}).
\end{equation}

The Fermi filling factors of independent Fermions
\begin{equation}
\label{np}
    n_p   =\frac{1}{\exp(2z_p)+1},\qquad
    z_p \equiv\frac{E_p}{2T},
\end{equation}
where $T$ is the temperature times $k_B$, are expressed by spectrum of
superconductor
\begin{equation}
\label{Ep}
    E_p =\sqrt{\xi_p^2+|\Delta_p|^2}, \qquad \xi_p=\varepsilon_p-E_F.
\end{equation}
Here we have to emphasize that for a model factorizable pairing potential
$V_{p,q}\propto \chi_p\chi_q$ the gap function is always separable as a
product of a temperature dependent function which can be associated with the
GL order parameter $Q(T)$ and a rigid temperature independent function of the
momentum $\chi_p$. The nontrivial results\cite{Pokrovskii:62} is that this
separation of the variables is asymptotically correct in the BCS weak-coupling
limit for an arbitrary kernel which is generally non factorizable. In fact, a
factorizable kernel is a fairly unnatural property which, however, can occur
if the pairing interaction is local, intra-atomic and located in a single atom
in the unit cell. This is the special case of the $s$-$d$ interaction at the
copper site(s) in the CuO$_2$ plane;\cite{Mishonov:03a} The separability
ansatz, though, shall be employed here to obtain a general interpolation
formula formally exact for factorizable kernels. We assume that the gap
anisotropy function $\chi_p$ is known, either as a result of solving the
general BCS equation at \Tc, inferred from experimental data processing, or
merely postulated within some model Hamiltonian, which is often the case for
the high-temperature and exotic superconductors.

With the above remarks, we will derive $C(T)$ for the separable gap
\begin{equation}\label{sepD}
    \Delta_p(T)=Q(T)\chi_p
\end{equation}
and a factorizable kernel.\cite{Markowitz:65}
We apply the ansatz (\ref{sepD}) to the BCS gap equation\cite{Bardeen:57}
\begin{equation}\label{BCS}
    \Delta_p(T)=\int\frac{d^Dq}{(2\pi\hbar)^D}V_{p,q}
    \frac{1-2n_q}{2E_q}\Delta_q(T),
\end{equation}
and use the convention that a positive sign of $V_{p,q}$ corresponds to
attraction of charge carriers and a negative potential energy of
interaction. Substituting here
\begin{align}
 \label{sepV}
    V_{p,q}&\approx G\chi_p\chi_q
\end{align}
and introducing $\eta\equiv|Q|^2$ we obtain a transcendental equation
for the temperature dependence of the gap $Q(T)$
\begin{align}\label{gA1}
    GA(\eta,T)&=1,\\
    A(\eta,T)&\equiv\overline{\left(\frac{\chi_p^2\tanh
    z_p}{2E_p}\right)}, \nn
\end{align}
where we have used the identity $1-2n_p=\tanh z_p$ and the coupling constant
is defined by $G\equiv 1/A(0,\Tc)$. Details on the derivation of the trial
function approximation \Eref{sepV} and the numerical solution of \Eref{gA1}
for $\Tc\ll\omega_D$ are given in Appendix~\ref{GapEqAppendix}.

For the specific heat of the superconducting phase per unit volume
divided by $k_B$ we have
\begin{equation}
\label{Cdiff}
    C(T)=T d_TS(\eta(T),T)=2\overline{E_p\,d_Tn_p}=C_\nu+C_\Delta,
\end{equation}
where $d_T=d/dT$. Here $C_\nu$ is the ``normal'' part of the specific heat
\begin{equation}
\label{Coverline}
    C_\nu(T)\equiv T(\partial_TS)_\eta=\frac{\pi^2}{3}\overline{
    g_c(z_p)},
\end{equation}
where
\begin{align}
    g_c(z)&\equiv\frac{6}{\pi^2}\frac{z^2}{\cosh^2z}, &
    \int_{-\infty}^{\,\infty} g_c(z) dz=1,
\end{align}
and $(\partial_T\ldots)_{\eta}$ is the temperature differentiation for
constant order parameter. For zero order parameter, $\eta=0$ at \Tc\ and
above, $C_\nu$ is just the specific heat of the normal phase
$C_N(T)=C_\nu(T,\eta=0)$.

Introducing
\begin{equation}
\label{GLa}
   \alpha({\eta},T)
   \equiv-(\partial_TA)_{\eta}
   = -(\partial_{\eta}S)_T
   = \frac{\overline{\chi_p^2\,g_a(z_p)}}{2T^2},
\end{equation}
where
\begin{align}
    \label{r_aDef}
    g_a(z)&\equiv\frac{1}{2\cosh^2z}, &
    \int_{-\infty}^{\,\infty} g_a(z)dz=1,
\end{align}
the other term of the specific heat
\begin{equation}
    C_\Delta\equiv T\partial_{\eta}
    S({\eta},T)d_T{\eta}(T)
\end{equation}
can be written as
\begin{equation}
    C_{\Delta}=\alpha({\eta},T)[- d_T{\eta}(T)]\,\theta(\Tc - T).
\end{equation}
\Eref{GLa} is actually a Maxwell-type equation
$\partial_{\eta}\partial_TF=\partial_T\partial_{\eta}F$, where $F$ is the free
energy: $S=-(\partial_TF)_{\eta}$, $A=-(\partial_{\eta}F)_T$;
cf. \Rref{Mishonov:02}.

Differentiating \Eref{gA1} we obtain $d A=0$ and
\begin{equation}
    -d_T{\eta}(T)=
    \frac{(\partial_TA)_{\eta}}{(\partial_{\eta}A)_T}
    \bigg|_{{\eta}(T)}=
    \frac{\alpha}{b}\bigg|_{{\eta}(T)},
\end{equation}
where the functions $\alpha$ and $b$ represent a generalization of the GL
coefficients for arbitrary temperature and order parameter
\begin{gather}
\label{GLb}
  b({\eta},T) \equiv-(\partial_{\eta}A)_T=\frac{7\zeta(3)}{16\pi^2T^3}
  \,\overline{\chi_p^4\,g_b(z_p)},\\
  g_b(z) \equiv\frac{\pi^2}{14\zeta(3)}\frac{1}{z^2}
    \left(\frac{\tanh z}{z}-\frac{1}{\cosh^2z}\right),\\
    \int_{-\infty}^{\,\infty} g_b(z) dz=1,
\end{gather}
and $\zeta$ is the Riemann zeta function. Then
\begin{equation}
    C_\Delta=T\frac{\alpha^2}{b}=\frac{4\pi^2}{7\zeta(3)}
    \frac{\left[\overline{\chi_p^2\,g_a(z_p)}\right]^2}
    {\overline{\chi_p^4\,g_b(z_p))}}\,\theta(\Tc - T)
\end{equation}
and
\begin{equation}
     \frac{C_\Delta}{C_{\nu}}=\frac{12}{7\zeta(3)}
     \frac{\left[\overline{\chi_p^2\,g_a(z_p)}\right]^2}
     {\overline{\chi_p^4\,g_b(z_p)}\;\;
     \overline{g_c(z_p)}}\,\theta(\Tc - T).
\end{equation}
The functions $g_i(z_p)$, $i=a,b,c$, introduced in
Refs.~\onlinecite{Mishonov:02} and \onlinecite{Mishonov:03b}, have sharp
maximum at the Fermi surface and in a good approximation we have
\begin{equation}
\label{FermiApprox}
  \overline{\chi_p^ng_i(z_p)} \approx 2T\nu_F\;
  \langle\chi_p^n\,r_i\left(y_p\right)\rangle,
  \qquad y_p\equiv\frac{\Delta_p}{2T},
\end{equation}
where
\begin{gather}
\label{rintegrals}
  r_i(y)\equiv\int_{-\infty}^{\,\infty}g_i(\sqrt{x^2+y^2}) dx,
  \qquad x=\frac{\xi_p}{2T},\\
  r_i(0)=1,\qquad r_i(\infty)=0,\qquad i=a,\;b,\;c.
  \nonumber
\end{gather}
We define averaging over the Fermi surface
\begin{equation}
  \label{FermiAveraging}
  \langle f_p\rangle=\frac{\overline{f_p\,\delta(\xi_p)}}
    {\nu_F}, \qquad
    \nu_F=\nu(E_F)=\overline{\delta(\xi_p)},
\end{equation}
where $\nu_F$ is the density of electron states per unit energy, volume and
spin at the Fermi level. In such a way we obtain
\begin{equation}
  C_{\nu}(T)=\frac{2}{3}\pi^2T\nu_F\;
  \langle r_{c}(y_p)\rangle
\end{equation}
and
\begin{equation}
  \frac{C_\Delta}{C_\nu}=\frac{12}{7\zeta(3)}
  \frac{\,\langle \chi_p^2\,r_{a}(y_p)\rangle ^2\mspace{10mu}\theta(\Tc-T)}
       {\langle \chi_p^4\,r_{b}(y_p)\rangle \mspace{10mu}\langle r_{c}(y_p)\rangle}.
\end{equation}
At \Tc, where the gap is small and $r_i(0)=1$ this formula gives
the Pokrovskii\cite{Pokrovskii:62} result for the reduced specific
heat jump
\begin{equation}
\label{CPokr}
  \frac{\Delta C}{C_N(\Tc)}=\frac{12}{7\zeta(3)}
  \frac{\langle \chi_p^2 \rangle ^2}
       {\langle \chi_p^4 \rangle}.
\end{equation}
For the GL coefficient \Eref{GLa} and \Eref{GLb} the approximation
(\ref{FermiApprox}) gives
\begin{align}
\label{aT}
  \alpha({\eta,T})&  =\frac{\nu_F}{T} \,\left\langle
  \chi_p^2\;r_a\left(\Delta_p/2T\right)\right\rangle,\\
  b({\eta,T}) &  =\frac{7\zeta(3)\nu_F}{8\pi^2T^2}
   \,\left\langle
  \chi_p^4\;r_b\left(\Delta_p/2T\right)\right\rangle.\nn
\end{align}
Then the specific heat takes the simple GL form for arbitrary
temperatures
\begin{equation}
\label{C(T)}
    C(T)=C_{\nu}(\eta,T)+T\frac{\alpha^2(\eta,T)}
           {b(\eta,T)}\,\theta(\Tc-T).
\end{equation}
Here, for the functions on the right-hand side we have substituted the thermal
equilibrium value of the order parameter $\eta(T)=|Q(T)|^2$, obtained from the
solution of \Eref{gA1}.  This BCS formula (\ref{C(T)}) is an example how good
the physical intuition was in the phenomenology of superconductivity.
According to the Gorter-Casimir\cite{Gorter:34} model the specific heat is a
sum of a ``normal'' part and another therm, governed by the temperature
dependence of the order parameter and having exactly the GL form. The
Gorter-Casimir two fluid model has very simple physical grounds. In the
self-consistent approximation, the entropy $S(T,\Delta(T))$ is a function of
the temperature and a temperature dependent order parameter $\Delta(T)$. The
temperature differentiation $C(T)=T(dS/dT)$ inevitably gives two terms in
\Eref{Cdiff}. According to the general idea by Landau,\cite{Ginzburg:50} the
order parameter is an adequate notion for description of second order phase
transitions, regardless of the concrete particle dynamics. The
$\epsilon$-expansion by Wilson and Fisher is only an ingenious realization of
the same Landau's idea when the influence of fluctuations is essential.

Again, at \Tc\  the general formulas \Eref{aT} give the Gor'kov and
Melik-Barkhudarov\cite{Gor'kov:63} result for the GL coefficients
\begin{equation}
  \alpha({0,\Tc})=\frac{\nu_F}{\Tc}\,\langle \chi_p^2\rangle ,\qquad
  b(0,\Tc)=\frac{7\zeta(3)\nu_F}{8\pi^2\Tc^2}\,\langle \chi_p^4\rangle.
\end{equation}
This result can be directly derived\cite{Mishonov:02} from the variational
free energy $F(\eta,T)$ of the superconductor which close to \Tc\ has the GL
form
\begin{equation}
F_{GL}(\eta,T)\approx\alpha({0,\Tc})\,(T-\Tc)\,|Q|^2
+\frac{1}{2}\,b(0,\Tc)\,|Q|^4.
\end{equation}
The simplest method to calculate the GL coefficients is to
differentiate\cite{Mishonov:02} the free energy after a $u$-$v$
transformations $F(\eta,T)=\overline{H}-TS$. Then
\begin{align}
\alpha(0,\Tc)=&(\partial_{\eta}F)_T(\eta=0,T=\Tc), \\
b(0,\Tc)=&(\partial_{\eta}^2F)_T(\eta=0,T=\Tc). \nonumber
\end{align}

\begin{figure}
\centering \includegraphics[width=8cm]{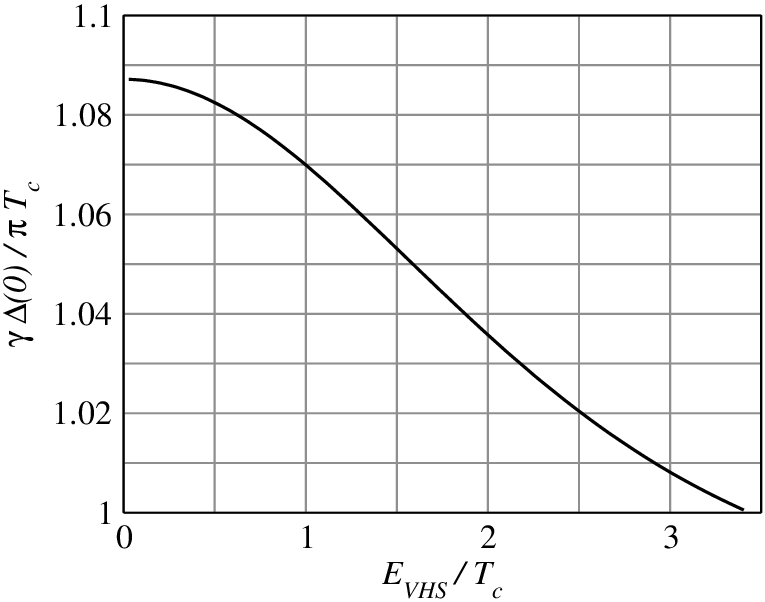}
\caption{$Z\equiv(2\Delta(0)/\Tc)/(2\pi/\gamma)$ vs $E_{\mathrm{VHS}}/\Tc$
computed for the model density of states \Eref{ln_VHS_DOS}. Note that 7\%
enhancement corresponds to $E_{\mathrm{VHS}}=\Tc$ and the maximum enhancement
is $\approx 9\%$. \label{fig-VHS}}
\end{figure}

If a Van Hove singularity (VHS) is close to the Fermi level the
formulas for GL coefficients are slightly modified\cite{Mishonov:03b}
\begin{align}
\alpha(0,\Tc)=&\frac{\langle\chi_p^2\rangle}{\Tc}
  \int_{-\infty}^{+\infty}\nu(E_F+2\Tc\,x)\,g_a(x)\,dx,\\
b(0,\Tc)=&\frac{7\zeta(3)\langle\chi_p^4\rangle}{8\pi^2\Tc^2}
  \int_{-\infty}^{+\infty}\nu(E_F+2\Tc\,x)\,g_b(x)\,dx, \nn \\
C_{\nu}(\Tc)=&\frac{2}{3}\pi^2\Tc
  \int_{-\infty}^{+\infty}\nu(E_F+2\Tc\,x)\,g_c(x)\,dx. \nn
\end{align}
Some important references on the influence of the VHS on the properties of
superconductors, and pioneering works on the two-band model are given in
\Rref{Mishonov:03b}. Let us evaluate the upper limit which can give a VHS. Let
us take 1D density of states $\nu(E)\propto 1/\sqrt{E-E_{\mathrm{VHS}}}$ and
$E_F=E_{\mathrm{VHS}}=0$; there is no doubt that this mathematical
illustration is unphysical. In this case we have for the reduced specific heat
jump $\Delta C/C_N(\Tc)$, \Eref{CPokr}, an additional factor
\begin{equation}
\frac{\left[\displaystyle\int_{0}^{\infty}g_a(\tilde{x}^2)d\tilde{x}\right]^2}
{\displaystyle\int_{0}^{\infty}g_c(\tilde{x}^2)d\tilde{x}
\displaystyle\int_{0}^{\infty}g_b(\tilde{x}^2)d\tilde{x}}=2.51, \quad
\tilde{x}\propto\sqrt{E}.
\end{equation}
Although this mathematical example is not realistic, it can be seen that the
VHS emulates qualitatively strong coupling corrections to the BCS theory: an
enhancement of $\Delta C/C_N(\Tc)$ and
$2\Delta_{\mathrm{max}}(0)/\Tc$. Another simulation of strong coupling effects
can be demonstrated by simple model density of states, corresponding to the
case of layered cuprates
\begin{equation}
\label{ln_VHS_DOS}
\nu(\xi)=1+k\ln\frac{1}{|\xi-E_{\mathrm{VHS}}|}.
\end{equation}
For illustration, we solve the equation
\begin{equation}
\int_{-\omega_D}^{\omega_D}
\frac{\tanh(\sqrt{\xi^2+\Delta^2(T)}/2T)}{2\sqrt{\xi^2+\Delta^2(T)}}
\, \nu(\xi)\,d\xi=G^{-1}
\end{equation}
taking $\omega_D=10$, $G=1/2$, and $k=10$. The
$Z\equiv(2\Delta(0)/\Tc)/(2\pi/\gamma)$ vs $E_{\mathrm{VHS}}/\Tc$ plot is
given in \Fref{fig-VHS}. It can be seen that 7\% enhancement corresponds to
$E_{\mathrm{VHS}}=\Tc$. Thus, the influence of the VHS on the specific heat is
much stronger than on the $\Delta(0)/\Tc$ ratio.

Let us also recall the general GL formula for the specific heat jump
at \Tc
\begin{equation}
\Delta C=\Tc\frac{\alpha^2(0,\Tc)}{b(0,\Tc)}.
\end{equation}
%
%============================= FIG. 1 ==================================
%=======================================================================

The two-band model provides probably the simplest possible illustration of the
derived formula for the specific heat; for pioneering references on the
two-band model see \Rref{Mishonov:03b}. The model is applicable with a
remarkable accuracy\cite{Bouquet:03} to MgB$_2$---a material which is in the
limelight in the physics of high-\Tc\ superconductivity over the past years.

For the normal specific heat we have
\begin{equation}
  C_{\nu}(T)=\frac{2}{3}\pi^2T\nu_F
  \left[c_1r_c(y_1)+c_2r_c(y_2)\right],
\end{equation}
where
\begin{equation}
y_1=\frac{\Delta_1}{2T},\qquad y_2=\frac{\Delta_2}{2T},
\qquad c_1+c_2=1,
\end{equation}
and $c_1\nu_F$ and $c_2\nu_F$ are the densities of states for the 2 bands of
the superconductor. Above \Tc\  or in the case of strong magnetic fields
$B>B_{c2}$ we have
\begin{equation}
\label{C_N}
C_N(T)=\frac{2}{3}\pi^2\nu_FT.
\end{equation}

As pointed out earlier, within the weak-coupling BCS approximation
Pokrovskii\cite{Pokrovskii:62} has proved the general separation of
the variables \Eref{sepD} which for a two-band superconductor
results in a weakly temperature dependent gap ratio
$\delta=\Delta_1/\Delta_2=\chi_1/\chi_2$. For MgB$_2$ determination of
the two gaps has been carried out by directional point-contact
spectroscopy\cite{Gonnelli:02} in single crystals.  One can see that
for model evaluations the temperature dependence of the gap ratio
could be neglected.

For the moments of the gap we have
\begin{equation}
 \left\langle
 \chi_p^n\,r_i\left(\frac{\Delta_p}{2T}\right)\right\rangle=
 \frac{c_1\delta^nr_i(y_1)+c_2r_i(y_2)}{(c_1\delta^2+c_2)^{n/2}},
 \quad i=a, b, c,
\end{equation}
where the normalization is irrelevant in further substitution in the
GL coefficients. Finally for the second, GL-order-parameter term of
the specific heat below the \Tc\ we obtain
\begin{equation}
\label{2bandC} C_\Delta(T)=\frac{8\pi^2}{7\zeta(3)}\nu_FT
\frac{\left[c_1\delta^2r_a(y_1)+c_2r_a(y_2)\right]^2}
{c_1\delta^4r_c(y_1)+c_2r_c(y_2)}.
\end{equation}
For the jump of the specific heat this formula reduces to the
Moskalenko\cite{Moskalenko:59} result
\begin{equation}
  \label{CJump}
  \frac{\Delta C}{C_N(\Tc)}
  =\frac{12}{7\zeta(3)}\,
  \frac{\left(c_1\chi_1^2+c_2\chi_2^2\right)^2}
  {c_1\chi_1^4+c_2\chi_2^4},
\end{equation}
which is, in fact, a special case of the Pokrovskii\cite{Pokrovskii:62}
formula \Eref{CPokr} applied to the two-band model. For application of the
two-band model to the specific heat of MgB$_2$ the reader is referred to
\Rref{Mishonov:03d}.

The analysis of the specific heat for MgB$_2$ gives perhaps the best
corroboration of the BCS results due to Pokrovskii\cite{Pokrovskii:62} and
Moskalenko.\cite{Moskalenko:59} Solving the Eliashberg equation and performing
first-principle calculations for the specific heat of MgB$_2$ Golubov~\etal\
[\Rref{Golubov:02a}, Fig.~3] derived 65$\%$ reduction of the specific heat
jump at \Tc. On the other hand, Eqs.~(\ref{CPokr}) (\ref{CJump}), using the
parameters from \Rref{Golubov:02a}, gives $\langle \chi^2\rangle^2/\langle
\chi^4\rangle=58\%$ reduction of the $\Delta C/C_N(\Tc)$ ratio. The 7$\%$
difference between those two estimates is in the range of the experimental
accuracy and the Eliashberg corrections to the BCS result is difficult to
extracted. Unfortunately, the groups solving the Eliashberg equation have not
compared their results to the classical results of the BCS theory for
anisotropic superconductors\cite{Pokrovskii:62} in order to analyze several
percent strong-coupling corrections to the specific heat jump for MgB$_2$.

In the single band case $c_1=1$ and \Eref{2bandC} gives a simple relation
between the specific heat and the BCS isotropic gap
\begin{equation}
\frac{C(T)}{C_N(T)} =r_{c}(y)+\frac{12}{7\zeta(3)}\,
\frac{r_a^2(y)}{r_b(y)},
\end{equation}
where $y(T)=\Delta(T)/2T.$
For anisotropic superconductors, functions of the gap have to be averaged
independently on the Fermi surface; this is the interpretation of the general
formulas \Eref{aT} and \Eref{C(T)}. Thus, we have the natural generalization
\begin{equation}
\label{C(T)_General} \frac{C(T)}{C_N(T)}=\langle r_c(y_p)\rangle +
\frac{12}{7\zeta(3)}\,\frac{\langle
\chi_{p}^{2}\,r_a(y_p)\rangle^2}{\langle
\chi_{p}^{4}\,r_b(y_p)\rangle},
\end{equation}
where $y_p(T)=\Delta_p(T)/2T=\chi_p\,Q(T)/2T.$

\begin{figure}
\centering
\includegraphics[width=8cm]{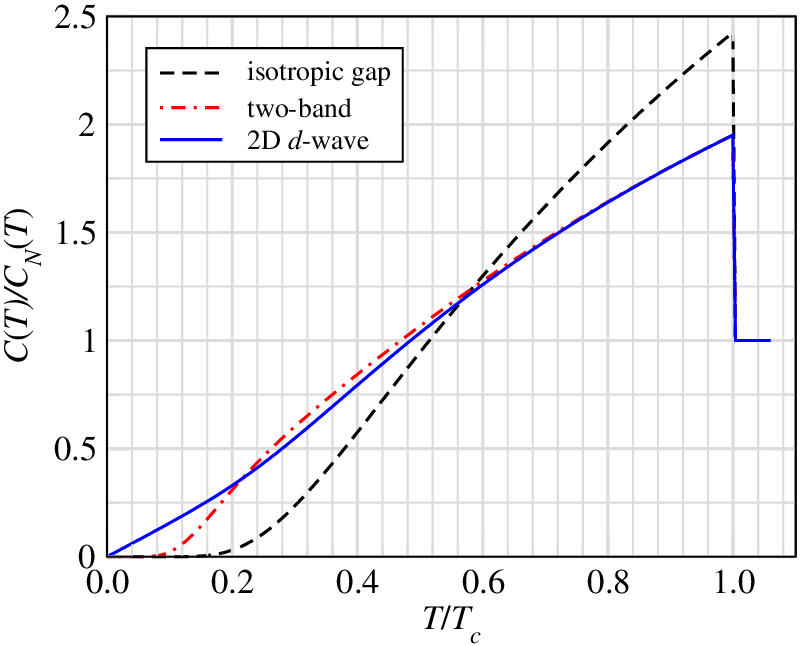}
\caption{Superconducting-to-normal specific heat ratio $C(T)/C_N(T)$ vs the
reduced temperature $t=T/\Tc$ according to \Eref{C(T)_General} computed for:
(i) an isotropic-gap BCS superconductor (dashed line), (ii) a two-band
superconductor $c_1=c_2=1/2$ with a gap ratio parameter $\delta=2.41$
(dash-doted line) and (iii) 2D $d$-wave superconductor $\chi_p=\cos 2\varphi$,
$\tan \varphi=p_x/p_y$ (solid line). Note that for $t>0.2$ two of the curves
would be experimentally indistinguishable. \label{fig-spec-heat}}
\end{figure}

For illustration, we now apply this general formula to three typical cases and
the results are shown in \Fref{fig-spec-heat}: (i) the isotropic-gap BCS model
$\chi_p=1$, familiar from a number of
textbooks;\cite{Lifshitz:81,Abrikosov:88,Abrikosov:75}
(ii) the two-dimensional (2D) $d$-wave superconductor
$\chi_p=\cos2\varphi$, $\tan\varphi=p_y/p_x$; and
(iii) a two-band superconductor $c_1=c_2=1/2$, for which the gap ratio
parameter is taken to reproduce the same reduced specific heat jump of
the $d$-wave superconductor ($\delta=\sqrt{3\pm\sqrt{8}}=2.41$ or
$0.41$).

The latter two models are often applied to analyze the behavior of CuO$_2$ or
MgB$_2$ superconductors. Note also the qualitative difference: for a $d$-wave
superconductor we have a quadratic specific heat at $T\ll \Tc$, whereas for a
two-band superconductor we have the exponential behavior $C(T)\propto
\exp(-\Delta_2/2T)$; see also \Fref{fig-spec-heat-MgB_2} below.

%============================= FIG. 2 ==================================
%=======================================================================

Consider now the low temperature behavior of the specific heat per
unit area for a 2D $d$-wave superconductors. Close to a node the gap
is proportional to the momentum component along the Fermi contour
$\Delta_p(0)\approx v_{\Delta} p_l$. The corresponding superfluid
velocity $v_{\Delta}$ is much smaller than the Fermi velocity $v_F$,
which parameterizes the dependence of the normal excitations energy
$\xi_p\approx v_F p_t$ as a function of the transversal to the Fermi
contour momentum component. For the ground state quasiparticle
spectrum we have $E_p\approx \sqrt{v_{\Delta}^2p_l^2+v_F^2p_t^2}$. It
is convenient to introduce the dimensionless variables
$q_1=v_{\Delta}p_l/2T$ and $q_2=v_Fp_t/2T$. In terms of the latter we
have for the element of the area in momentum space
\begin{equation}
4\frac{dp_ldp_t}{(2\pi\hbar)^2}=4\frac{(2T)^2}{v_{\Delta}v_F}\,\frac{2\pi
q\,dq}{(2\pi\hbar)^2}=\frac{2EdE}{\pi\hbar^2v_{\Delta}v_F},
\end{equation}
where $q=\sqrt{q_1^2+q_2^2}=z_p=E_p/2T$, and for axial symmetric functions we
can use polar coordinates; cf. \Rref{Hussey:02}. Here we have taken into
account 4 nodal points. In such a way \Eref{Coverline} gives
\begin{equation}
C_{\nu}(T\ll \Tc) =\frac{16}{\pi\hbar^2}\frac{T^2}{v_{\Delta}v_F}
\int_{0}^{\infty}\frac{q^3\,dq}{\cosh^2q}
\approx6.89\frac{T^2}{\hbar^2v_{\Delta}v_F},
\end{equation}
where we used $18\zeta(3)/\pi\approx 6.89$; cf. \Rref{Hussey:02},
Eq.~(2.9). This result together with \Eref{C_N} gives for the
superconducting-to-normal specific heat ratio
\begin{equation}
\frac{C_{\nu}}{C_N}(t\ll 1)
=1.047\,\frac{\Tc}{\hbar^2\nu_F\,v_{\Delta}v_F}\,t,
\end{equation}
where $t=T/\Tc$ is the reduced temperature. The penetration depth has a
similar linear low temperature behavior for $d$-wave superconductors.

%============================= FIG. 3 ==================================
\begin{figure}
\centering
\includegraphics[width=8cm]{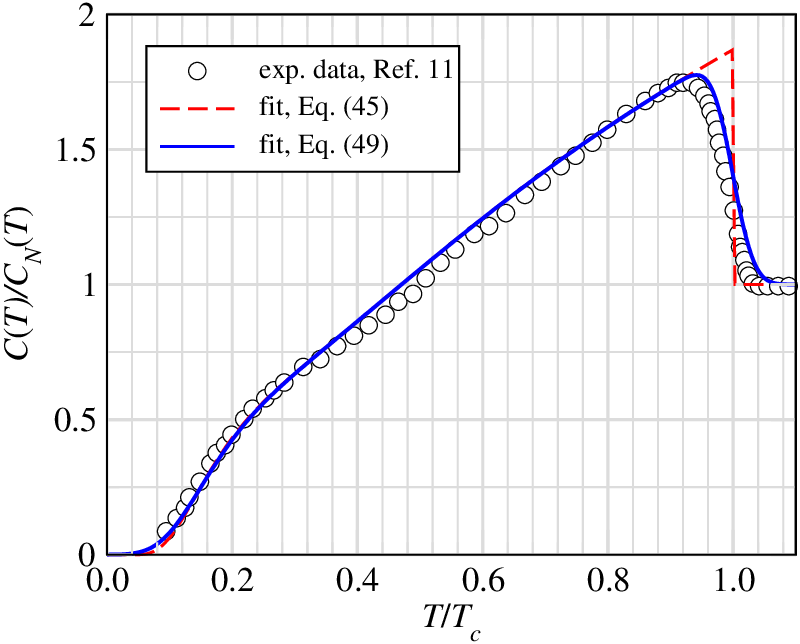}
\caption{Comparison between the superconducting-to-normal specific heat ratio
$C(T)/C_N(T)$; the theoretical curve is computed following \Rref{Mishonov:03d}
with $c_1=0.49$, $\delta=2.9$ (solid line) and the experimental data for
MgB$_2$ are taken from \Rref{Bouquet:03} (circles). The theoretical curve is
convoluted with a Gaussian kernel \Eref{Cconv}, chosen to fit best the
experimental data ($\Delta t=0.027$). The experimental data\cite{Bouquet:03}
are digitized from \Rref{Mishonov:03d}, Fig.~3.
\label{fig-spec-heat-MgB_2}}
\end{figure}
%=======================================================================

Very often fluctuations of stoichiometry and crystal defects make the theory
of homogeneous crystal inapplicable close to the critical region. Let
$\Tc(\mathbf{r})$ be a weakly fluctuating Gaussian field of the space vector
$\mathbf{r}$. Hence, the simplest possible empirical model is to apply a
Gaussian kernel to the theoretically calculated curve. Then for the heat
capacity we have
\begin{equation}
\label{Cconv} C(t)=\int_{-\infty}^{+\infty} C_{\mathrm{theor}}(t)
\, \exp\left\{-\frac{(t-t')^2}{2(\Delta t)^2}\right\}
\frac{dt'}{\Delta t \sqrt{2\pi}}.
\end{equation}
The philosophy of applying the convolution technique to all theoretical curves
with singularities was advocated in the book by Migdal.\cite{Migdal:75} Such
an empirically smeared curve with $\Delta t=0.027$ describes better the
experimental data for MgB$_2$ close to \Tc; $\Tc\Delta t \approx 1.1$ K,
$B_{c2}(0)=2.5$ T and $B_{c2}(0)\Delta t=750$ G. The result is depicted at
\Fref{fig-spec-heat-MgB_2}, where the smeared theoretical curve is compared
with the experimental data.\cite{Bouquet:03} In order to achieve a good fit of
the theory to the experimental data we have treated $c_1$ and $\delta$ as
fitting parameters (cf. Refs.~\onlinecite{Bouquet:03},
\onlinecite{Mishonov:03d}, \onlinecite{Fisher:03} and
\onlinecite{Choi:03}). The values used $c_1=0.49$ and $\delta=2.9$ are
slightly different from the set of parameters used latter for computing the
penetration depth, but are still in agreement with different spectroscopic
evaluations.  In order to reach the analogous quality of the fit of $C(T)$ for
cuprates we have to take into account simultaneously the gap anisotropy and
the VHS in the general expressions \Eref{GLa} and \Eref{GLb}.

An analogous to \Eref{Cconv} smearing of the fluctuation magnetization above
\Tc\ reads
\begin{align}
&M(B,T-\Tc)\\
&=\int M_{\mathrm{theor}}(B,T-\Tc')\,
\exp\left\{-\frac{(\Tc'-\Tc)^2}{2(\Tc\Delta t)^2}\right\}\,
\frac{d\Tc'}{\sqrt{2\pi}\Tc\Delta t}. \nn
\end{align}
However, for big fluctuations of \Tc\ we have to take into account the
appearance of superconducting domains. Such a precise investigation of
fluctuations in the magnetization of Nb and Sn in the past led to the
discovery of twinning plane superconductivity.  For analytical GL results for
twinning plane superconductivity see \Rref{Mishonov:90}.

Here we wish to emphasize that a large body of experimental data for
$B_{c2}(T)$ are strongly influenced by the disorder. It is imperative to cut
off a region of width $\Tc\Delta t$ or $B_{c2}(0)\Delta t$ close to
$B_{c2}(\Tc)$ if we wish to determine $B_{c2}(T)$ by extrapolation of
properties from the superconducting phase or fluctuation behavior of the
normal phase. Various spurious curvatures of $B_{c2}(T)$ have been reported
merely as a result of disorder of the crystals.

%=======================================================================
\section{Electrodynamic behavior}
%=======================================================================

An analysis of the London penetration depth tensor, similar to that carried
out by Kogan in \Rref{Kogan:02}, gives
\begin{equation}
\label{LambdaKogan} (\lambda^{-2}(T))_{\alpha\beta}=
\frac{e^2}{\varepsilon_0 c^2}2\nu_F \langle r_d(y_p)
v_{\alpha}v_{\beta} \rangle, \quad \alpha,\beta=x,y,z,
\end{equation}
where
\begin{equation}
 \mathbf{v}_p=\frac{\partial\varepsilon_p}{\partial\mathbf{p}}, \qquad
 m_p^{-1}=\frac{\partial\mathbf{v}_p}{\partial\mathbf{p}}
 =\frac{\partial^2\varepsilon_p}{\partial\mathbf{p}^2}
\end{equation}
are the band velocity and effective mass and
\begin{gather}
\label{ii2d2}
r_d(y) \equiv(y/\pi)^2\sum_{n=0}^{\infty}
\left[(y/\pi)^2+\left(n+\frac{1}{2}\right)^2\right]^{-3/2},\\
 r_d(y) \approx 7\zeta(3)(y/\pi)^2\ll1,\qquad
 r_d(\infty)=1.
\nn
\end{gather}
For comparison, the conductivity tensor of the normal phase in
$\tau_p$-approximation reads
\begin{equation}
\sigma_{\alpha\beta}= 2\nu_F e^2 \langle \tau_p \,
v_{\alpha}v_{\beta} \rangle.
\end{equation}
For the penetration depths along the principal crystal axes we have in
the two-band model
\begin{equation}
\lambda_{\alpha}^{-2}(T)=
   \lambda_{\alpha,1}^{-2}(0)\,r_d(y_1)
  +\lambda_{\alpha,2}^{-2}(0)\,r_d(y_2),
\end{equation}
where for uniaxial crystals like MgB$_2$ there are only 4 constants:
$\lambda_{x,1}(0)=\lambda_{y,1}(0)$, $\lambda_{x,2}(0)=\lambda_{y,2}(0)$,
$\lambda_{z,1}(0)$ and $\lambda_{z,2}(0)$. These can be obtained from electron
band calculations,\cite{Mishonov:03c}
\begin{align}
\label{lamda_ab_2_band}
(\lambda^{-2}(T))_{\alpha\beta} &=\frac{e^2}{\varepsilon_0 c^2}
2\nu_F \sum_{b=1,2}c_b r_d(\frac{\sst\Delta_b(T)}{\sst 2T}) \langle
v_{\alpha}v_{\beta} \rangle_b\\
 &
=\sum_{b=1,2}(\lambda_b^{-2}(0))_{\alpha\beta}\,r_d(\frac{\sst\Delta_b(T)}{\sst
  2T}),
\nn
\end{align}
where the band index $b$ labels the leaf of the Fermi surface over which the
averaging of the electron velocities is carried out. For a discussion and
details see the review by Kogan and Bud'ko.\cite{Kogan:02}
There is a natural ``Eliashbergization'' of this result (cf.
Refs.~\onlinecite{Golubov:02a}, \onlinecite{Choi:03},
\onlinecite{Golubov:02b,Zehetmayer:03,Waelte:04}):
\begin{align}
  r_d(\frac{\sst\Delta_p}{\sst 2T}) & =\sum_{n=0}^{\infty}
  \frac{2\pi T\Delta_p^2}{\left(\Delta_p^2+\omega_n^2\right)^{3/2}} \\
  &\longrightarrow \sum_{n=0}^{\infty}
  \frac{2\pi T\tilde{\Delta}_p^2}
  {\left[\tilde{\Delta}_p^2(\omega_n)+\tilde{\omega}_{n,p}^2\right]^{3/2}},
  \nonumber
\end{align}
where $\omega_n=(2n+1)\,\pi\,T$ are the Matsubara frequencies,
$\tilde{\omega}_{n,p}=Z_p(\omega_n)\,\omega_n$,
$\tilde{\Delta}_p(\omega_n)=Z_p(\omega_n)\,\Delta_p(\omega_n)$ and
$Z_p(\omega_n)$ is the normalization factor. Analogous expressions can be
worked out for the specific heat.

For a heuristic consideration of the result by Kogan\cite{Kogan:02} at $T=0$
see \Rref{Mishonov:03c}. At $T=0$ the Fermi surface is shifted as a rigid
object in the momentum space under the influence of electromagnetic
field. This shift of all conduction electrons explains why for the penetration
depth the influence of VHS is less essential than the influence on the heat
capacity. The increase of the kinetic energy of all conduction electrons is
actually the increase of the Gibbs free energy density $\Delta
G=\frac{1}{2\varepsilon_0c^2}\lambda^2j^2$. At finite temperatures the number
of superfluid electrons is $r_d(\Delta_p/2T)$ times smaller.

The penetration depths at $T=0$ can be also expressed by the optical masses
and the Hall constant of the normal metal at high magnetic field
\begin{gather}
 (\lambda^{-2}(0))_{\alpha\beta}=\frac{e}{\varepsilon_0 c^2}
   \frac{1}{\mathcal{R}_{\infty}}(m^{-1})_{\alpha\beta},\\
 \frac{1}{\mathcal{R}_{\infty}}
   = 2e\int_{\varepsilon_p<E_F}\frac{d^3p}{(2\pi\hbar)^3},\nn\\
m^{-1} = \frac{\displaystyle\int_{\varepsilon_p<E_F}
         \frac{d^3p}{(2\pi\hbar)^3}m^{-1}_p}
{\displaystyle\int_{\varepsilon_p<E_F}\frac{d^3p}{(2\pi\hbar)^3}}
 = \frac{\displaystyle\oint_{\varepsilon_p=E_F}
   \frac{dS_p}{(2\pi\hbar)^3v_p} \mathbf{v}_p \otimes \mathbf{v}_p}
   {\displaystyle\int_{\varepsilon_p<E_F}\frac{d^3p}{(2\pi\hbar)^3}},
\nn
\end{gather}
the last equation being a consequence of the Gauss theorem
$\int_{\varepsilon_p<E_F}
d^3p\frac{\partial}{\partial\mathbf{p}}=\oint_{\varepsilon_p=E_F}
d\mathbf{S}_p$, where $d\mathbf{S}_p$ is the element of the Fermi surface
oriented along the outward normal.
For an extensive discussion on galvanomagnetic properties of normal metals and
inclusion of hole pockets with volume density $n_h$ for
$\mathcal{R}_{\infty}^{-1}=e(n_e-n_h)$ see the textbook by Lifshitz and
Pitaevskii\cite{Lifshitz:79} or the monograph by Lifshitz, Azbel and
Kaganov.\cite{Lifshitz:71} The Bernoulli effect can be easily observed in
almost compensated superconductors for which $n_e\approx n_h$ and the Hall
constant is bigger.

In the superconducting phase the Hall constant $\mathcal{R_{\infty}}$
can be determined by the Bernoulli potential
\begin{equation}
\Delta\varphi=-\mathcal{R_{\infty}}\frac{1}{2\varepsilon_0 c^2}
\lambda^2(T) j^2;
\end{equation}
generalization for the anisotropic case can be obtained by the obvious
replacement $\lambda^2j^2\rightarrow
j_{\alpha}\lambda_{\alpha\beta}^2j_{\beta}$. Here we suppose that $j\ll
j_c(T)$, $j_c$ being the critical current. If the magnetic field $\mathbf{B}$
is parallel to the surface of a bulk superconductor this formula gives
\begin{equation}
\Delta\varphi=-\mathcal{R_{\infty}}\frac{B^2}{2\mu_0}.
\end{equation}
All charge carriers interact with the electric potential $\varphi$,
but only the superfluid part $\propto r_d(\Delta_p/2T)$ creates
kinetic energy. The constancy of the electrochemical potential in the
superconductor gives the change of the electric potential, i.e., the
Bernoulli effect. For the temperature dependent condensation energy
$\Delta G=-B_c^2(T)/2\mu_0$ the corresponding change of the electric
potential is given by
\begin{equation}
\Delta\varphi=\mathcal{R_{\infty}}\frac{B_c^2(T)}{2\mu_0}.
\end{equation}
For complete determination of the Hall constant
$\mathcal{R_{\infty}}$, the penetration depth $\lambda(T)$ and the
optical mass of conduction electrons in a clean superconductor
[cf. \Rref{Mishonov:03c}, Eq.~(20)],
\begin{equation}
m=\frac{e\lambda^2(0)}{\varepsilon_0c^2\mathcal{R_{\infty}}},
\end{equation}
we have to investigate the Bernoulli effect for thin,
$d_{\mathrm{film}}\ll\lambda(T)$, and thick,
$d_{\mathrm{film}}\gg\lambda(T)$, superconducting films of the same
material. $M_{\mathrm{cp}}\equiv 2m$ can be called effective mass of
the Cooper pairs; this parameter can be significantly increased by
disorder.

For the temperature dependence of the electrochemical potential of the
normal phase we have [\Rref{Lifshitz:71}, Eq.~(12.16)]
\begin{equation}
e\Delta\varphi=\frac{\pi^2}{6}\,\frac{\nu'(E_F)}{\nu(E_F)}\,T^2.
\end{equation}
Close to a VHS the influence of the energy derivative of the density
of states can be significant and measurable.

%============================= FIG. 4 ==================================
\begin{figure}[t]
\centering
\includegraphics[width=8cm]{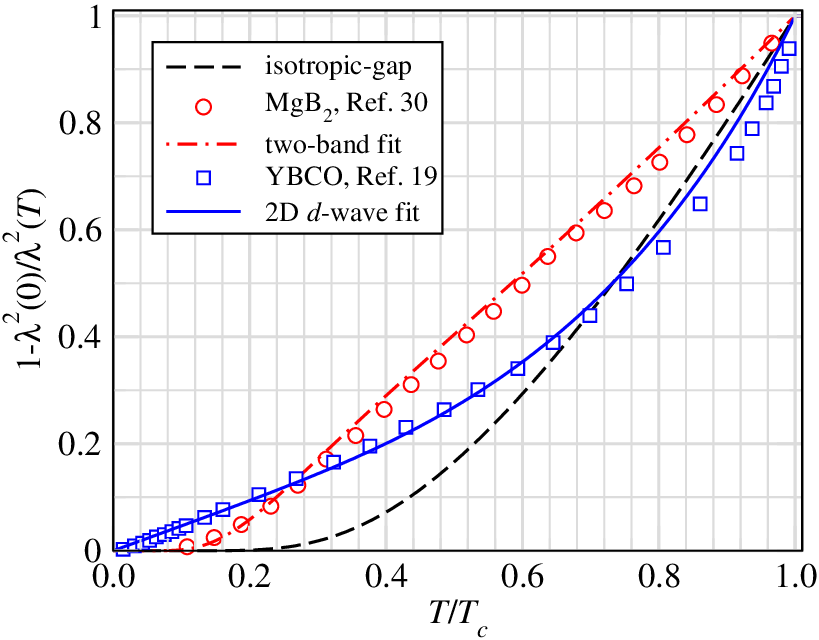}
\caption{In-plane normal fluid density $1-\lambda^2(0)/\lambda^2(T)$ vs
reduced temperature $t=T/\Tc$ computed for 3 cases: (i) isotropic-gap BCS
superconductor (dashed line), (ii) two-band superconductor MgB$_2$ with
parameters $c_1=0.59$, $\delta=7.1/2.8$ (dash-doted line) and (iii) 2D
$d$-wave superconductor (solid line). The experimental points for
YBa$_2$Cu$_3$O$_{7-\delta}$ (squares) are digitized from \Rref{Hussey:02} and
the corresponding theoretical 2D $d$-wave curve is calculated according to
\Eref{rho_S_remorm} with renormalization factor $Z=1.4$. Some experimental
points for MgB$_2$ (circles) are digitized from \Rref{Carringtone:03}, Fig.~9;
for details see the original work.
\label{fig-normal-fluid-density}}
\end{figure}
%=======================================================================

The entropy and specific heat related to the volume density of the free energy
of superconducting condensation $B_c^2(T)/2\mu_0$ can be determined by
electric capacitor measurements, applying surface temperature
oscillations. For discussions of possible experimental setups see
\Rref{Mishonov:03c} and references therein.

It is a matter of technical calculations to verify the identity
\begin{align}
 (y/\pi)^2 \sum_{n=0}^{\infty} &
 \left[(y/\pi)^2+\left(n+\frac{1}{2}\right)^2\right]^{-3/2}\\
 &+\int_{-\infty}^{+\infty}\frac{dx}{2\cosh^2\sqrt{x^2+y^2}}=1, \nn
\end{align}
which transcribes into the form
\begin{equation}
\label{rad}
r_a(y) + r_d(y) = 1.
\end{equation}
In such a way the \emph{electrodynamic behavior} of a superconductor can be
expressed in terms of the functions, defined for description of its
\emph{thermodynamic behavior}. Using Eqs.~(\ref{rad}) and (\ref{LambdaKogan})
we obtain
\begin{equation}
\label{rho_normal}
\rho_{\mathrm{N}}(T)=
 1-\frac{(\lambda^{-2}(T))_{\alpha\beta}}{(\lambda^{-2}(0))_{\alpha\beta}}
 =\frac{\left<
 r_a(\displaystyle\frac{\sst \Delta_p}{\sst 2T})v_{\alpha}v_{\beta}\right>}{\langle
 v_{\alpha}v_{\beta}\rangle}.
\end{equation}
Within the framework of London electrodynamics
$\rho_{\mathrm{N}}(T)=1-\lambda^2(0)/\lambda^2(T)$ is the normal fluid
density, and $\rho_{\mathrm{S}}(T)=\lambda^2(0)/\lambda^2(T)$ is the
superfluid one, having total charge density
$\rho_{\mathrm{S}}(T)/\mathcal{R}_{\infty}$. For a two-band superconductor,
Eqs. (\ref{lamda_ab_2_band}) and (\ref{rho_normal}) give for the penetration
depth along the principal crystal axes
\begin{align}
\rho_{\mathrm{S}}(T)=\frac{\lambda_{\alpha}^2(0)}{\lambda_{\alpha}^2(T)}
=&\!\sum_{b=1,2}\!w_{\alpha,b}r_d(\frac{\Delta_b(T)}{2T}), \quad
w_{\alpha,b}=c_b\!\frac{\langle v_{\alpha}^2\rangle_b}
{\langle v_{\alpha}^2\rangle}, \nonumber \\
w_{{\alpha},1}+w_{{\alpha},2}=&1, \quad \langle
v_{\alpha}^2\rangle=c_1\langle v_{\alpha}^2\rangle_1+ c_2\langle
v_{\alpha}^2\rangle_2.
\end{align}

For a set of parameters see the review by Kogan and Bud'ko.\cite{Kogan:02} We
take $\delta=7.1/2.8$ according to the spectroscopic data;\cite{Mishonov:03d,
Daghero:03} see also the point contact spectroscopy data in
\Rref{Samuely:03}. In \Fref{fig-normal-fluid-density} we compare our
theoretical calculation with the experimental data for $\lambda(T)$ by
Carrington and Manzano.\cite{Carringtone:03} Here we take $c_1=0.59$ which
gives $w_{a,1}\approx w_{a,2}\approx 0.5$.

%============================= FIG. 5 ==================================
\begin{figure}[t]
\centering \includegraphics[width=8cm]{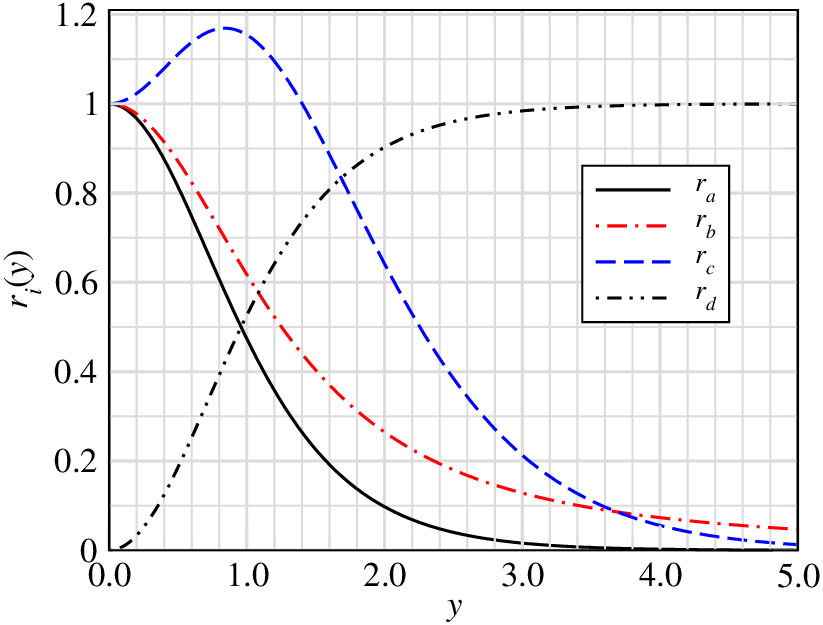}
\caption{Plot of the $r_i(y)$ functions ($i=a,b,c,d$).
\label{fig-r-functions}}
\end{figure}
%=======================================================================

\begin{figure}[t]
\centering
\includegraphics[width=8cm]{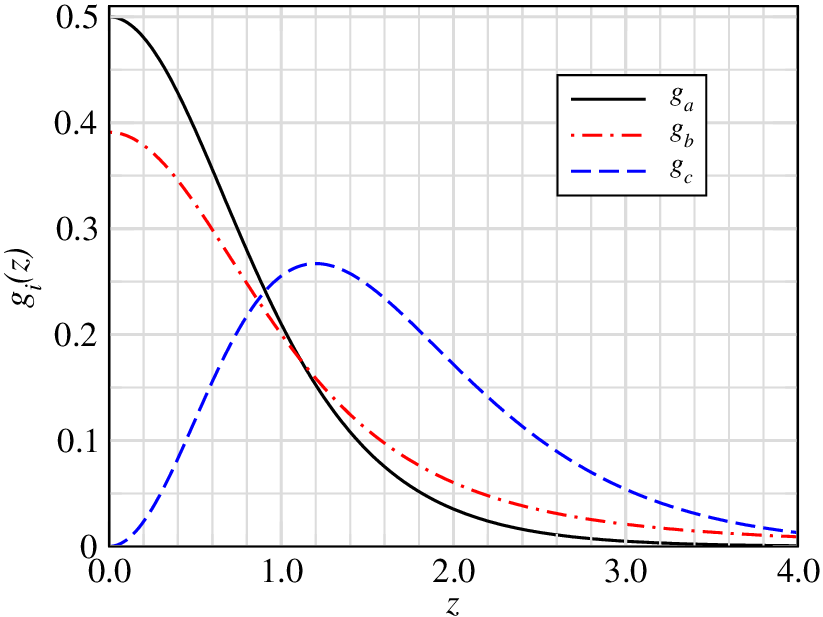}
\caption{Plot of the $g_i(z)$ functions ($i=a,b,c$).
\label{fig-g-functions}}
\end{figure}

The functions $r_i(y)$ for $i=a,\,b,\,c,\,d$ can be easily programmed for the
purposes of experimental data processing. The graphs of $r_i(y)$ and the
corresponding $g_i(z)$ functions are given in \Fref{fig-r-functions} and
\Fref{fig-g-functions}. The temperature dependence of the penetration depth
$\lambda(T)$ is also programmed for isotropic-gap, two-band and model 2D
$d$-wave superconductor. In the 2D $d$-wave case the theoretical result is
compared with the experimental data\cite{Hussey:02} for
YBa$_2$Cu$_3$O$_{7-\delta}$, which is also depicted at the figure. The linear
dependence of $1-\lambda^2(0)/\lambda^2(T)$ at low temperatures for
YBa$_2$Cu$_3$O$_{7-\delta}$ is discussed in \Rref{Hussey:02}, Eq.~(2.10). For
a 2D $d$-wave superconductor the general formula \Eref{LambdaKogan} gives
\begin{equation}
\label{rho_S_remorm}
\rho_{\mathrm{S}}(T)=\frac{\lambda^2(0)}{\lambda^2(T)}=
\int_{0}^{2\pi}
r_d(Z\frac{\sst \Delta_{\mathrm{max}}(T)}{\sst 2T}\cos2\varphi)
\frac{d\varphi}{2\pi},
\end{equation}
where the temperature dependence of the order parameter is described in
Appendix~\ref{GapEqAppendix}. We are using an oversimplified model for cuprate
superconductivity for which are neglected (i) the anisotropy of the Fermi
velocity $v_F(\mathbf{p})$ along the Fermi contour; (ii) higher harmonics of
the gap function $\Delta_p$ along the Fermi contour and (iii) the influence of
VHS of the density of states slightly below the Fermi level.
For comparison between Angle Resolved Photoemission Spectroscopy (ARPES) data
and a lattice model for high-\Tc\ spectrum see \Rref{Mishonov:03a}, Fig. 3.

Let us assume now that the order parameter for YBa$_2$Cu$_3$O$_{7-\delta}$ is
$Z$-times higher than the BCS prediction. This could be due to the influence
of VHS or, which is more important, strong coupling effects. Inserting here
$Z=1.4$ we can see that such a renormalization well describes the temperature
dependence of the penetration depth in the whole temperature interval.
Finally, we have a good working BCS-like formula. In fact, significantly
higher $\Delta_{\mathrm{max}}(0)/\Tc$ than BCS prediction is in agreement with
the ARPES data.
%============================= FIG. 6 ==================================
%=======================================================================

%=======================================================================
\section{The case for
  S\lowercase{r}$_{\mathbf{2}}$R\lowercase{u}O$_{\mathbf{4}}$}
%=======================================================================

\begin{figure}
\centering
\includegraphics[width=8cm]{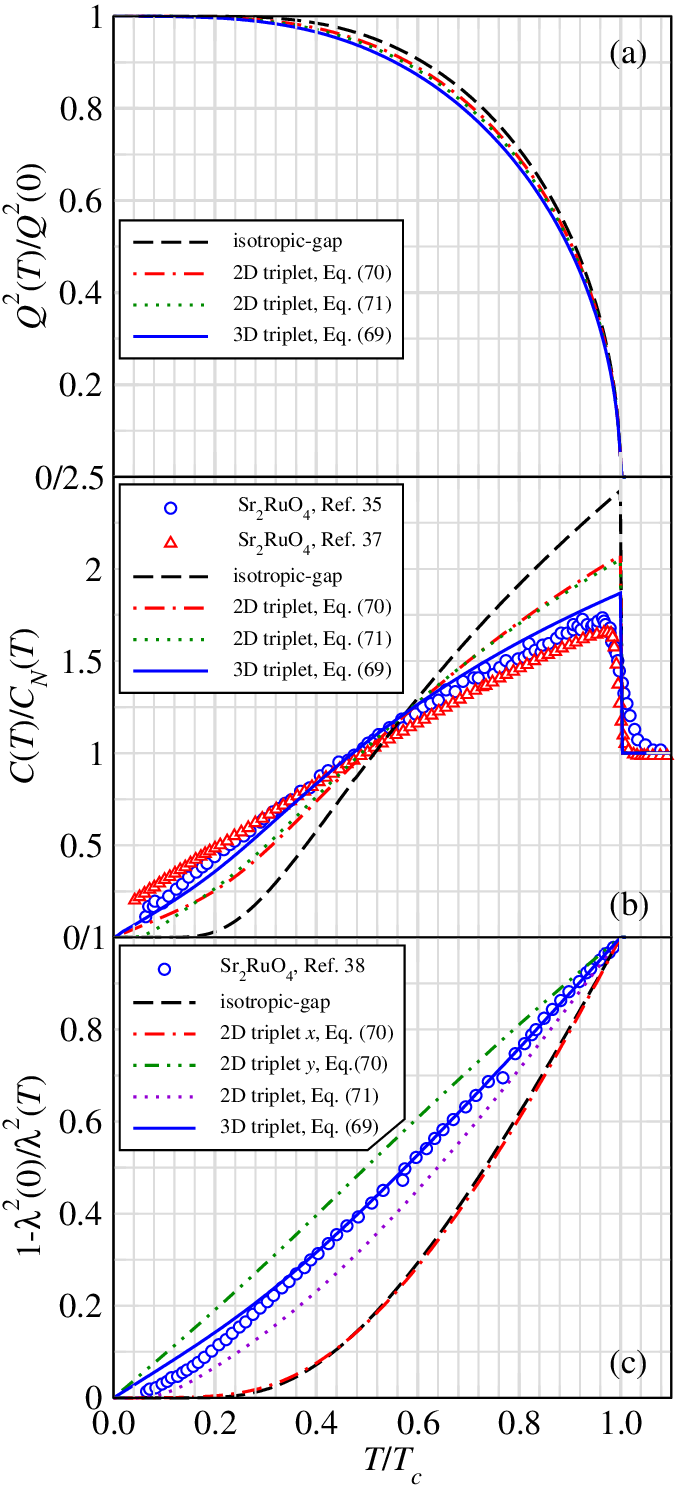}
\caption{Sr$_2$RuO$_4$. (a) Reduced order parameter for the Zhitomirsky and
Rice model \Eref{SRO_Zhitomirsky} (solid line), the 2D vertical line nodes
model \Eref{SRO_Nishizaki} (dot-dashed line), and for the 2D model by
Deguchi~\etal\ \Eref{gap_deguchi} (doted line). (b) Specific heat ratio
$C(T)/C_N(T)$ for the Zhitomirsky and Rice model (solid line), the 2D vertical
line nodes model (dash-dotted line), and for the Deguchi~\etal\ model (doted
line). The experimental points (circles) from \Rref{Nishizaki:99} are
digitized from \Rref{Zhitomirsky:01}, Fig.~1. (c) Normal fluid density
$1-\lambda^2(0)/\lambda^2(T)$ corresponding to the gap anisotropy models
(\ref{SRO_Zhitomirsky})--(\ref{gap_deguchi}).  The experimental points
(circles) from \Rref{Bonalde:00} are digitized from \Rref{Dahm:00}, Fig.~2. We
should note that the model with vertical line nodes predicts spontaneous
breaking of the symmetry of the penetration depth in the $ab$-plane.
\label{fig-SRO}}
\end{figure}

Our approach is aslo applicable to the triplet superconductor Sr$_2$RuO$_4$;
for a review see \Rref{Mackenzie:03}. We adopt the promising gap anisotropy
model by Zhitomirsky and Rice,\cite{Zhitomirsky:01} which gives
$E_p=\sqrt{\smash[b]{\xi_p^2+|\Delta_p|^2}}$, with
\begin{equation} \label{SRO_Zhitomirsky}
|\Delta_p|^2 \propto \left[
\sin^2\frac{p_xa}{2\hbar}\cos^2\frac{p_ya}{2\hbar}
 + \cos^2\frac{p_xa}{2\hbar}\sin^2\frac{p_ya}{2\hbar} \right]
\cos^2\frac{p_zc}{2\hbar},
\end{equation}
where $p_xa/\hbar$, $p_ya/\hbar$, $p_zc/\hbar \in (0,2\pi)$. For the Fermi
surface we take a simple cylinder
$\varepsilon_p\approx\varepsilon(\sqrt{\smash[b]{p_x^2+p_y^2}})$ with radius
$p_Fa/\hbar\approx 0.93\,\pi$. Our calculations are depicted in
\Fref{fig-SRO}. In this model calculation we have taken into account only one
band responsible for superconductivity. Although it is not \textit{a priori}
clear how ``good'' is this assumption, our curve reproduces the theoretical
curve by Zhitomirsky and Rice\cite{Zhitomirsky:01} and passes close to the
experimental points by NishiZaki~\etal\cite{Nishizaki:99} This promising
success encouraged us to present the theoretical prediction for the
penetration depth calculated from \Eref{rho_normal}. According to the
conclusions by Zhitomirsky and Rice\cite{Zhitomirsky:01} their model with
horizontal line nodes (see also \Rref{Litak:04}) describes the experimental
data better than a model with vertical line nodes. For illustration, in
\Fref{fig-SRO} we present also our calculations for a simple 2D vertical line
nodes model with gap anisotropy function
\begin{equation}
\chi_p \propto \sin\left(\frac{p_xa}{2\hbar}\right). \label{SRO_Nishizaki}
\end{equation}
Similar model was studied by NishiZaki~\etal;\cite{Nishizaki:99} see also
Fig.~26 in the review by Mackenzie and Maeno.\cite{Mackenzie:03}

From aesthetic point of view our preferences are for the recent model for the
gap anisotropy by Deguchi~\etal\cite{Deguchi:04}
\begin{equation}
\label{gap_deguchi}
|\Delta_p|^2\propto \sin^2(p_x a/\hbar) + \sin^2(p_ya/\hbar).
\end{equation}
Such type of anisotropy can be derived in the framework of
quasi-two-dimensional exchange models for perovskite superconductivity of the
type of the considered for CuO$_2$ plane in \Rref{Mishonov:03a}. The
theoretical prediction corresponding to \Eref{gap_deguchi} is also illustrated
in Fig.~\ref{fig-SRO} together with the experimental data by
Deguchi~\etal\cite{Deguchi:04}

%============================= FIG. 7 ==================================
%=======================================================================

%=======================================================================
\section{Discussion and conclusions}
%=======================================================================

Let us discuss now the specific heat. We have shown that for factorizable
kernels\cite{Markowitz:65} the specific heat can be represented as a sum of a
``normal'' component $C_\nu(T)$ and a term dependent on the order parameter
$C_\Delta(T)$, which has the same form as in the GL theory. There is one
detail that is worth focusing on: for the $s$-$d$ model for high-\Tc\
superconductivity\cite{Mishonov:03a} the kernel is indeed separable because
the contact interaction is localized in a single atom in the lattice unit
cell. One should only substitute the spectrum of the superconductor at $T<\Tc$
in the known expression for the GL coefficients from classical work of Gor'kov
and Melik-Barkhudarov.\cite{Gor'kov:63} The final expression for the specific
heat is a generalization of the result of Pokrovskii.\cite{Pokrovskii:62} The
derived formulas can be easily programmed for fitting the experimental data of
anisotropic superconductors. For the jump of the specific heat at the critical
temperature $\Delta C|_{\Tc}=C_{\Delta}(\Tc^-)$ general consideration has
already been given in \Rref{Mishonov:02}. The derived formula is not exact,
but interpolates between the correct low-temperature behavior and the result
by Pokrovskii\cite{Pokrovskii:62} for the specific heat jump at $\Tc.$ That is
why we believe that our interpolation formula \Eref{C(T)} can be useful for
preliminary analysis of the experimental data for the specific heat in
superconductors; for experimental data processing the accuracy could be
comparable, e.g., with the accuracy of the Debye formula for the phonon heat
capacity.

We illustrated our formulas for $C(T)$ and $\lambda(T)$ for isotropic-gap BCS
model and three of the best investigated anisotropic-gap superconductors
YBa$_2$Cu$_3$O$_{7-\delta}$, Sr$_2$RuO$_4$ and MgB$_2$. The nature of
superconductivity for those superconductors is completely different:
high-\Tc\  and low-\Tc, phonon- and exchange-mediated, singlet and triplet
Cooper pairs. In all those cases the derived formulas work with an acceptable
accuracy; in some cases we have even quantitative agreement and for high-\Tc\ 
cuprates we have shown what the BCS analysis can give. We conclude that the
statistical properties of the superconductors [thermodynamic $C(T)$ and
kinetic $\lambda(T)$] are determined mainly by the gap anisotropy,
irrespective of the underlying pairing mechanism, and the approximative weak
coupling separation of variables\cite{Pokrovskii:62} $\Delta_p(T)=Q(T)\chi_p$
is an adequate approach. It is worth applying the derived formulas for $C(T)$
and $\lambda(T)$ for every new superconductor. Often after the synthesis of a
new superconductor single crystals are not available and only the data for heat
capacity $C(T)$ can help the theory to distinguish between different models
for the gap anisotropy even before detailed spectroscopic investigation is
performed.

\begin{acknowledgments}
One of the authors (TM) is thankful to F.~Bouquet, A.~Carrington, K.~Deguchi,
L.P.~Gor'kov, N.~E.~Hussey, A.~Junod, V.G.~Kogan, Y.~Maeno, V.A.~Moskalenko,
V.L.~Pokrovsky, A.~Rigamonti, Y.~Wang, O.~Dolgov and O.K.~Andersen for
clarifying correspondence related to their papers. The authors are thankful to
L.A. Atanasova and T.I. Valchev for the cooperation in the initial stage of
the present research. This work was partially supported by the Scientific Fund
of Sofia University under Contract 57/2004.
\end{acknowledgments}

\begin{appendix}

\section{Order parameter equation for anisotropic-gap superconductors}
\label{GapEqAppendix}

Following \Rref{Pokrovskii:62}, let us scrutinize the derivation of and the
solution to \Eref{gA1}. The gap anisotropy function will have non-zero values
only in a narrow region near the Fermi surface
\begin{equation}
\chi_p=\chi_p\,\theta(\omega_D-|\xi_p|),\qquad \Tc\ll\omega_D\ll E_F.
\end{equation}
Later, the differential volume in the momentum space can be separated to Fermi
surface element $dS$ and a normal element $dp_t$
\begin{equation}
d^Dp=dp_t\,dS=\frac{d\varepsilon}{v_F}\,dS,\qquad
v_F(\mathbf{p})=\left|\frac{\partial
\varepsilon_p}{\partial\mathbf{p}}\right|.
\end{equation}
Returning to \Eref{gA1} we have
\begin{equation}
\frac{G}{(2\pi\hbar)^D} \oint\!\!\int\frac{\chi_p^2}{2E_p}\tanh(z_p)
 \, \theta(\omega_D-|\xi_p|) \, \frac{d\varepsilon\,dS}{v_F} \,=1,
\end{equation}
where $\oint$ denotes integration over the Fermi surface. With the account of
the energy cutoff $\omega_D$ the last reads
\begin{equation}
\label{gA1DeriveTmp1}
 \frac{G}{(2\pi\hbar)^D}\oint\frac{dS}{v_F}\,\chi_p^2\,
 \int_{0}^{\omega_D}\frac{\tanh(\sqrt{\smash[b]{\xi^2+\Delta_p^2}}/2T)}
{\sqrt{\xi^2+\Delta_p^2}}\,d\xi=1.
\end{equation}
According to \Eref{FermiAveraging} we have for the density of states
\begin{align}
\nu_F=\overline{\delta(\xi_p)}=&\frac{1}{(2\pi\hbar)^D}
\int\delta(\varepsilon-E_F)d\varepsilon\,\frac{dS}{v_F}\\
=&\frac{1}{(2\pi\hbar)^D}\oint\frac{dS}{v_F}. \nonumber
\end{align}
Similarly, the averaging over the Fermi surface can be represented as a
surface integral
\begin{equation}
\langle f(p)\rangle
=\frac{1}{\nu_F}\oint\frac{dS}{(2\pi\hbar)^D\,v_F}f(p).
\end{equation}
In these notation \Eref{gA1DeriveTmp1} reads
\begin{equation}
\label{gA1Derive1}
 \left\langle\chi_p^2\int\limits_{0}^{\omega_D}\frac{\tanh(\sqrt{\smash[b]{\xi^2+\Delta_p^2}}/2T)}
 {\sqrt{\xi^2+\Delta_p^2}}\,d\xi\right\rangle=\frac{1}{G\nu_F}=
 \frac{1}{\lambda_{\mathrm{BCS}}},
\end{equation}
where $\lambda_{\mathrm{BCS}}\equiv G\nu_F$ is the dimensionless BCS coupling
constant.

At $T=\Tc$, where $\Delta_p=0$ and $E_p=|\xi_p|$, substituting $x=\xi/2T$ we
obtain
\begin{equation}
\label{gA1DeriveT_c} \langle\chi_p^2\rangle\int_{0}^{M}\frac{\tanh
x}{x}\,dx=\frac{1}{\lambda_\mathrm{BCS}}, \qquad
M=\frac{\omega_D}{2\Tc}\gg1.
\end{equation}
Now the identity
\begin{equation}
\int_{0}^{M}\frac{\tanh x}{x}\,dx=\ln\left(\frac{4\gamma}{\pi}M\right)
\end{equation}
gives
\begin{equation}
\label{gA1Derive2}
 \Tc=2\omega_D\,\frac{\gamma}{\pi}\,
 \exp\left(-\frac{1}{\langle\chi_p^2\rangle\,\lambda_{\mathrm{BCS}}}\right).
\end{equation}
Analogously, at $T=0$ we have
\begin{equation}
\label{gA1DeriveTmp2} \left\langle \chi_p^2
\int_{0}^{\omega_D}\frac{d\xi}{\sqrt{\xi^2+\Delta_p^2(0)}}\right\rangle
 = \frac{1}{\lambda_{\mathrm{BCS}}}.
\end{equation}
Then taking into account that $\omega_D\gg\Delta_p(0)$ we have
\begin{equation}
\label{gA1DeriveTmp3}
 \int\limits_{0}^{\omega_D}\frac{d\xi}{\sqrt{\xi^2+\Delta_p^2}}
 = \ln\left( \frac{\omega_D}{|\Delta_p|} +
   \sqrt{1+\frac{\omega_D^2}{|\Delta_p|^2}}\,\,\right)\approx
   \ln\frac{2\omega_D}{|\Delta_p|}.
\end{equation}
As we will see later, it is convenient to modify the normalization of the
order parameter and gap anisotropy function:
\begin{equation}
\tilde{\chi}_p=\frac{\chi_p}{\chi_{\mathrm{av}}},\quad
\tilde{Q}=Q\,\chi_{\mathrm{av}},\quad \chi_{\mathrm{av}}\equiv\exp\left\{
\frac{\langle \chi_p^2\,\ln|\chi_p| \rangle}{\langle
\chi_p^2\rangle} \right\}.
\end{equation}
The renormalizing multiplier $\chi_{\mathrm{av}}$ is chosen in order for the
renormalized gap anisotropy function to obey the relation
\begin{equation}
\label{chi_tilde_norm}
\langle \tilde{\chi}_p^2\,\ln\tilde{\chi}_p^2\rangle=0.
\end{equation}
For the two-band model this gives
\begin{equation}
\chi_{\mathrm{av}} =\chi_1^{c_1\chi_1^2} \chi_2^{c_2\chi_2^2},
\end{equation}
and one can easily verify that
\begin{equation}
c_1\tilde{\chi}_1^2\,\ln|\tilde{\chi}_1|+c_2\tilde{\chi}_2^2\,
\ln|\tilde{\chi}_2|=0.
\end{equation}
Similarly, using
\begin{equation}
\int_0^{\pi/2}\cos^2\varphi\,\ln|\cos\varphi|\,d\varphi
 = \frac{\pi}{8}\,\ln(\mathrm{e}/4)
\end{equation}
we obtain for a 2D $d$-wave superconductor
\begin{gather}
\tilde{\chi}_p(\varphi)=\frac{2}{\sqrt{\mathrm{e}}}\,\cos2\varphi,\\
\int_0^{2\pi}\tilde{\chi}_p^2(\varphi)\,
\ln|\tilde{\chi}_p(\varphi)|\,d\varphi=0.\nonumber
\end{gather}
Using the approximation (\ref{gA1DeriveTmp3}) with a renormalized order
parameter and gap anisotropy function, from \Eref{gA1DeriveTmp2} we derive
\begin{equation}
\label{gA1Derive3}
\tilde{Q}(0)=2\omega_D\,\exp\left(-\frac{1}{\langle\chi_p^2\rangle\lambda_{\mathrm{BCS}}}\right).   
\end{equation}
This equation together with (\ref{gA1Derive2}) gives the well-known BCS
relation for the renormalized order parameter for anisotropic
superconductors\cite{Pokrovskii:62}
\begin{equation}
\label{gA1Derive4}
\frac{2\tilde{Q}(0)}{\Tc}=\frac{2\pi}{\gamma}\approx3.53.
\end{equation}
We assume that the density of states $\nu(E)$ is almost constant in the energy
interval $E_F\pm 2\Tc$.

The renormalization does not change the gap
$\Delta_p(T)=Q\,\chi_p=\tilde{Q}\,\tilde{\chi_p}$, but in a sense
$\tilde{Q}(T)$ is the ``true'' BCS gap for an anisotropic superconductor. For
$T=0$ the BCS model gives for $d$-wave superconductors
$\Delta_p(0)=\Delta_{\mathrm{max}}\cos2\varphi$, where
\begin{equation}
\frac{2\Delta_{\mathrm{max}}}{\Tc}=\frac{2\pi}{\gamma}\,\frac{2}{\sqrt{\mathrm{e}}}\approx4.28.
\end{equation}
However, for cuprates we have to take into account the influence of Van Hove
singularity and strong coupling correlations. As we fitted from the
temperature dependence of the penetration depth for
YBa$_2$Cu$_3$O$_{7-\delta}$ we have 40\% bigger gap
$\Delta_{\mathrm{max}}=Z\Delta_{\mathrm{max}}^{(\mathrm{BCS})}$ and
$2\Delta_{\mathrm{max}}/\Tc \approx 6.0$. In such a way the thermodynamic
behavior is in agreement with the spectroscopic data. This is a good hint in
favor of the Landau-Bogoliubov quasiparticle picture applied to high-\Tc\
cuprates. For MgB$_2$ taking $c_1=0.44$ and $\Delta_1(0)=7.1$ meV and
$\Delta_2(0)=2.8$ we obtain $\tilde{\chi}_1\approx 1.17$ and
$\tilde{\chi}_2\approx 0.46$. Then $\tilde{Q}(0) =\Delta_1(0)/\tilde{\chi}_1
=\Delta_2(0)/\tilde{\chi}_2 \approx 6.08$ meV $=70.6$~K. For the critical
temperature $\Tc=39$~K we obtain $2\tilde{Q}(0)/\Tc \approx 3.62$ which agrees
with the BCS ratio (\ref{gA1Derive4}) within $3\%$ accuracy as found in
\Rref{Mishonov:03d}

For arbitrary temperatures using the identity
\begin{equation}
\tanh\frac{x}{2} = 1 - \frac{2}{\mathrm{e}^x+1}
\end{equation}
\Eref{gA1Derive1} reads
\begin{align}
\label{gA1DeriveTmp4}
&\left\langle\chi_p^2\int_{0}^{\omega_D}
  \frac{d\xi}{\sqrt{\xi^2+\Delta_p^2(T)}}\right\rangle
 -\frac{1}{\lambda_{\mathrm{BCS}}}\\
=& 2\left\langle\chi_p^2\int\limits_{0}^{\omega_D} \frac{d\xi}{
\sqrt{\xi^2+\Delta_p^2(T)} \left[
\exp\left(\frac{\sqrt{\xi^2+\Delta_p^2(T)}}{T}\right)+1 \right] }
\right\rangle. \nonumber
\end{align}
Substituting here $1/\lambda_{\mathrm{BCS}}$ from \Eref{gA1DeriveTmp2} and
taking into account the $\omega_D\gg |\Delta_p(0)|$ approximation,
\Eref{gA1DeriveTmp3}, we obtain the Pokrovskii equation
\begin{equation}
\label{GapMainEq} q:=\exp\left\{-\frac{\left\langle
\chi_{p}^{2}F(2y_p
)\right\rangle}{\langle\chi_p^2\rangle}\right\}, \quad
 2y_p = \frac{\pi}{\gamma}\,\frac{\chi_p}{\chi_{\mathrm{av}}}\,\frac{q}{t}
      = \frac{\Delta_p}{T},
\end{equation}
where
\begin{equation}
q(t)=\frac{\Delta_p(T)}{\Delta_p(0)}= \frac{Q(T)}{Q(0)}
=\frac{\tilde{Q}(T)}{\tilde{Q}(0)}
\end{equation}
is the reduced order parameter $0 \leq q \leq 1$ as a function of the reduced
temperature $t=T/\Tc$. In physical variables Pokrovskii\cite{Pokrovskii:62}
equation reads
\begin{equation}
\ln\frac{\Delta_p(T)}{\Delta_p(0)} + \left\langle \chi_p^2\, F(\Delta_p(T)/T)
\right\rangle_p=0.
\end{equation}
The function $F(x)$ associated with the right-hand side of
\Eref{gA1DeriveTmp4} is defined by an integral, for which we have one integral
and two different summation formulas, convenient for small and large
arguments\cite{Rumer:72}
\begin{align}\label{Ffunc}
F(x)
 &\equiv \int_{-\infty}^{\infty}\frac{du}{\sqrt{u^2+x^2}
  \left[\exp\left(\sqrt{u^2+x^2}\right)+1\right]}\\
 &= 2\int_{0}^{\infty}\frac{du}{\exp(x\cosh u)+1}\nonumber\\
 &= \ln\frac{\pi}{\gamma x}
    + 2\pi\sum_{l=1}^{\infty}\left[\frac{1}{(2l-1)\pi}
    -\frac{1}{\sqrt{x^2+(2l-1)^2\pi^2}}\right] \nonumber\\
 &= -2 \sum_{n=1}^{\infty}(-1)^{n}K_0(nx), \nonumber
\end{align}
where for large arguments we have the approximate formula
\begin{equation}
2K_0(x\gg 1) \approx \sqrt{\frac{2\pi}{x}}\,\mathrm{e}^{-x}\!\!
 \left(
        1 -\frac{1}{8x} +\frac{9}{128x^2} -\frac{225}{3972x^3}
 \right).
\end{equation}
Physically, here $x=\Delta/T$, $u=\xi/T$ and the upper integration bound
$\omega_D/T$ has been replaced by $\infty$. For this function we have the
approximate formulas
\begin{align}
\label{FLowApprox} F(x\ll1)&\approx \ln\frac{\pi}{\gamma
x}+\frac{7}{8\pi^2}\,\zeta(3)\,x^2,\\
\label{FHighApprox} F(x\gg1)&\approx 2K_0(x).
\end{align}
The Euler constant is $\gamma=\mathrm{e}^C\approx 1.781072418$ and
$\zeta(3)\approx 1.202056903$, where $\zeta$ is the Riemann zeta function. A
plot of the function $F(x)$ is shown in \Fref{fig-F-function}. In
Appendix~\ref{CppAppendix} a simple \textsc{C++} code for numerical evaluation
of this function is provided. The use of the \texttt{limes()} function is
optional; it increases the accuracy, but slows down the computation. For fast
calculations we have to take only several terms of the expansions
\Eref{Ffunc}.
The colon (:) in \Eref{GapMainEq} represents an iterative assignment in which
we use the initial approximation $q=1$.

%============================= FIG. 8 ==================================
\begin{figure}[t]
\centering
\includegraphics[width=8cm]{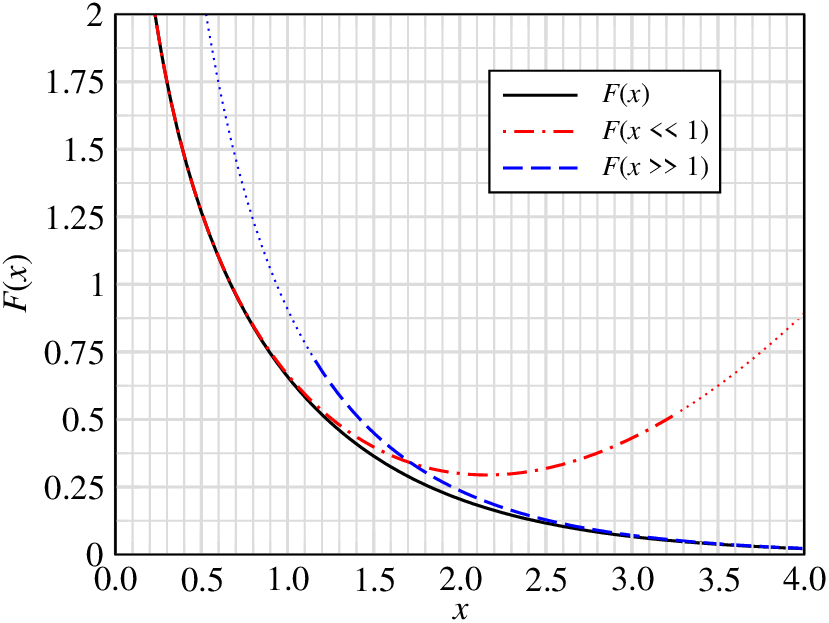}
\caption{Plot of the $F(x)$ function. The approximations to $F(x)$ for $x\ll1$
  and $x\gg1$ are given by Eqs.~(\ref{FLowApprox}) and (\ref{FHighApprox}),
  respectively.\label{fig-F-function}}
\end{figure}
%=======================================================================

The BCS order parameter equation \Eref{GapMainEq} is not specific for the
physics of superconductivity. Recently, Abrikisov\cite{Abrikosov:03} has
derived the same equation for the temperature dependence of the amplitude of
spin density waves in cuprates.

\begin{figure}[t]
\centering
\includegraphics[width=8cm]{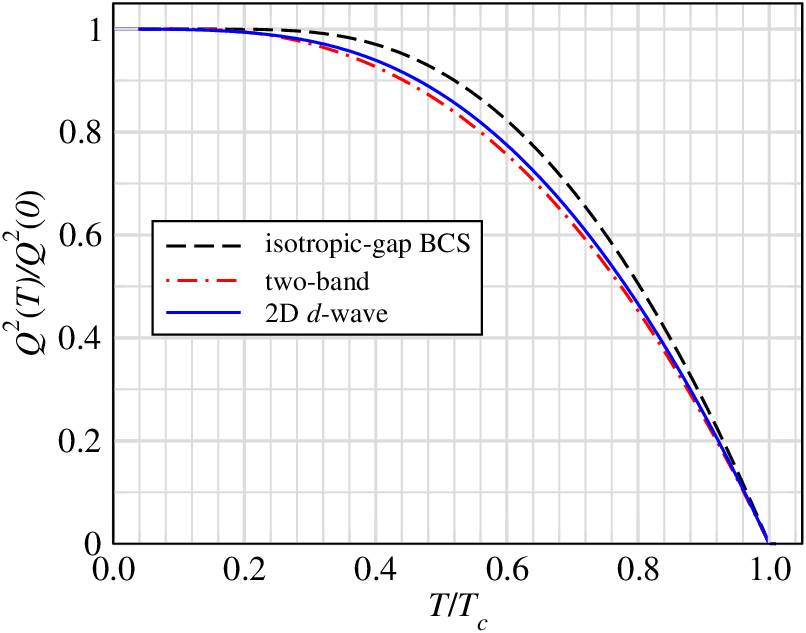}
\caption{Squared reduced order parameter $Q^2(T)/Q^2(0)$ vs reduced
temperature $t=T/\Tc.$ For the two-band model, the $c_1$ and $\delta$
parameters are chosen so as to simulating a $d$-wave CuO$_2:$ $c_1=1/2$,
$\delta=2.41$. \label{fig-gap-square}}
\end{figure}

\begin{figure}
\centering
\includegraphics[width=8cm]{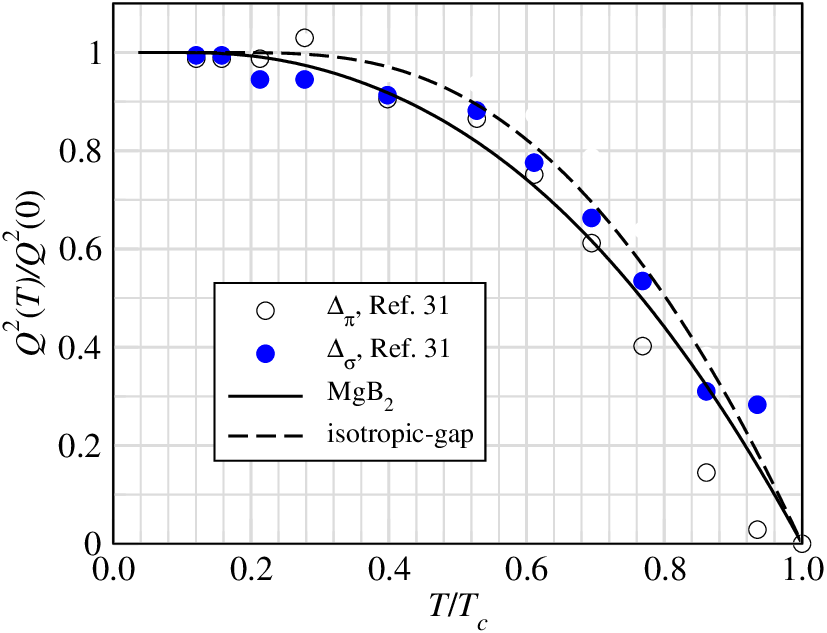}
\caption{Squared reduced order parameter $Q^2(T)/Q^2(0)$ vs the reduced
temperature $t=T/\Tc$ for MgB$_2$ (solid line) with $c_1=0.44$,
$\delta=7.1/2.8$. The experimental points for MgB$_2$ (circles) are digitized
from \Rref{Daghero:03}.
\label{fig-gap-MgB_2}}
\end{figure}

For 2D $d$-wave superconductors the Pokrovskii equation (\ref{GapMainEq})
reads
\begin{equation}
\ln q =-\int\limits_{0}^{2\pi} 2\cos^2(2\varphi) \,
F\left(\frac{2\pi}{\gamma\sqrt{\smash[b]{\mathrm{e}}}}\cos(2\varphi)
\frac{q}{t}\right)\,\frac{d\varphi}{2\pi}.
\end{equation}
The numerical solution for the squared reduced order parameter $q^2(t)$ is
shown in \Fref{fig-gap-square}. The linear dependence near the critical
temperature $t=1$ corresponds to the GL approximation. In \Fref{fig-gap-MgB_2}
the squared reduced order parameter for MgB$_2$ (two-band model with
$c_1=0.44$, $\delta=7.1/2.8$) is compared with the experimental data from
\Rref{Daghero:03}.

%============================= FIG. 9 ==================================
%=======================================================================

As a last problem, let us derive the factorizable kernel (\ref{sepV}) as a
result from the BCS equation~(\ref{BCS}). For $\omega_D\ll E_F$, \Eref{BCS}
reads
\begin{align}
\label{fac1} \Delta_q(T) &= \oint_{\mathrm{FS}}V_{q,p}\,\Delta_p
\int\limits_{0}^{\omega_D} \frac{\tanh(E_p/2T)}{E_p}\, d\xi_p
 \frac{dS_p}{(2\pi\hbar)^Dv_F} \nonumber\\
 &= \nu_F \left\langle V_{q,p}\Delta_p \int_{0}^{\omega_D}
 \frac{\tanh(E_p/2T)}{E_p}\, d\xi_p \right\rangle_p.
\end{align}
At $T=\Tc$ [cf. Eqs.~(\ref{gA1DeriveT_c})--(\ref{gA1Derive2})] this formula
gives
\begin{equation}
\label{fac2} \Delta_q(\Tc)\approx\nu_F
 \ln\left(\frac{2\gamma\omega_D}{\pi\Tc}\right)
 \langle V_{q,p}\Delta_p \rangle_p.
\end{equation}
Let us also mention the dimensions of the variables. Since the integration
$\int\cdots\frac{d^3p}{(2\pi\hbar)^3}$ has a dimension of $1/$volume, then
$V_{q,p}$, being a Fourier component of potential energy, has dimension of
energy$\times$volume. For example, the Coulomb potential $e^2/r$ has dimension
of energy and its Fourier transformation
\begin{equation}
\frac{4\pi e^2}{k^2}=
\int\frac{e^2}{r}\,\mathrm{e}^{-\,i\,\mathbf{k}\cdot\mathbf{r}}\,
d^3r
\end{equation}
has dimension of energy$\times$volume. The same holds for the contact
attraction in the BCS model potential $V(\mathbf{r})=-G\delta(\mathbf{r})$
having a constant Fourier component $-G$. The density of states $\nu_F$ has
dimension of (energy$\times$volume)$^{-1}$, $\Delta_p$ and $E_p$ have
dimension of energy and the Fermi surface averaging brackets
$\langle\ldots\rangle$ represent a dimensionless operation.

Let the dimensionless parameter $V_0$ denotes the maximum eigenvalue of the
problem
\begin{equation}\label{fac4_eigen}
 \langle V_{q,p}\,\chi_p \rangle_p=V_0\,\chi_q,
\end{equation}
and $\chi_p$ is the corresponding eigenvector, with normalization
$\langle\chi_p^2\rangle=1$. The comparison of \Eref{fac4_eigen} and
\Eref{fac2} gives
\begin{equation}
\Tc=\frac{2\gamma\omega_D}{\pi}\exp\left(-\frac{1}{\nu_F V_0}\right),
\end{equation}
which is identified with \Eref{gA1Derive2} and we obtain
\begin{equation}
 G=V_0=\frac{\langle\chi_qV_{q,p}\chi_p\rangle_{q,p}}
 {\langle\chi_p^2\rangle_p}.
\end{equation}
As the maximal eigenvalue is sought, one can apply in this case the Krilov
iterations
\begin{equation}
\chi_q^{(n+1)}\propto\langle V_{q,p}\chi_p^{(n)}\rangle_p, \qquad
\langle(\chi_p^{(n+1)})^2\rangle=1,
\end{equation}
starting from some solution-like trial vector $\chi_p^{(0)}$. Then the gap
anisotropy function $\chi_p$ is just the limit of the Krilov iterations
$\chi_p^{(\infty)}$.

%============================= FIG. 10 =================================
%=======================================================================

For $T=0$, the gap equation (\ref{fac1}) gives
\begin{equation}
\label{fac5} \Delta_p(0)=\left< V_{q,p}
 \ln\left(\frac{2\omega_D}{\tilde{Q}(0)|\tilde{\chi}_p|}\right)
 \Delta_p \right>_p.
\end{equation}
Within the weak-coupling BCS approximation, in the integrant
\begin{equation}\label{fac6}
  \ln\left(\frac{2\omega_D}{|\Delta_p(0)|}\right)
 =\ln\left(\frac{2\omega_D}{\tilde{Q}(0)}\right) - \ln|\tilde{\chi}_p|
\end{equation}
the first term is much bigger than the second one. For details we refer to the
original work by Pokrovsky,\cite{Pokrovskii:62} but roughly speaking
$\ln[2\omega_D/\Delta_p(0)]\approx\mathrm{const}\gg 1$. Within the latter
approximation for $\Delta_p(0)$ we obtain again the same eigenvalue problem
and this constitutes the proof that the momentum dependence of the gap is
rigid. Hence we derive the separation of the variables $\Delta_p(T)\approx
Q(T)\chi_p$. When the term $\ln|\chi_p|$ in \Eref{fac6} is small it can be
treated perturbatively, and according to the normalization
\Eref{chi_tilde_norm} its influence diminish. The properties of this
approximative separation of the variables can be simulated by a factorizable
kernel
\begin{equation}
\label{fac7} V_{q,p} =\sum_{n} V_n\Psi_q^{(n)}\Psi_p^{(n)}
\rightarrow V_0\chi_q\chi_p,
\end{equation}
where $V_n$ are the eigenvalues and $\Psi_p^{(n)}$ are the corresponding
eigenvectors of the problem
\begin{equation}
\langle V_{q,p}\,\Psi_p^{(n)} \rangle_p=V_n\,\Psi_q^{(n)}, \qquad
\langle|\Psi_p^{(n)}|^2\rangle=1.
\end{equation}
In other words, the factorizable approximation, \Eref{fac7} and \Eref{sepV},
works well when the influence of smaller eigenvalues is small.

Generally speaking, the separability ansatz is a low-\Tc\ approximation; \Tc\
should be much smaller than all other energy parameters: energy cutoff, Debye
frequency for phonon superconductors, exchange integrals for exchange mediated
superconductivity, the Fermi energy and the bandwidths. Room temperature
superconductivity is not yet discovered, but the good message is that we have
still a simple approximation acceptably working for all superconductors.
For theoretical models the accuracy of the separable approximation can be
easily probed when investigating the angle between the order parameter at
different temperatures, e.g.,
\begin{equation}
\arccos \frac{\langle\Delta_p^*(T)\Delta_p(\Tc)\rangle}
{\sqrt{\langle|\Delta_p(T)|^2\rangle\langle|\Delta_p(\Tc)|^2\rangle}}
\ll 1,
\end{equation}
or
\begin{equation}
\arccos \frac{\overline{\Delta_p^*(T)\Delta_p(\Tc)}}
{\sqrt{\overline{|\Delta_p(T)|^2}\,\,\overline{|\Delta_p(\Tc)|^2}}}
\ll 1.
\end{equation}
Those angles are just zero at \Tc\ and the expressions for the specific heat
jump and the GL coefficients is correct. Only for $T\to 0$ some small
deviations can be observed, but in that case one can treat $\chi_p$ as a trial
function in a variational approach.

The performed analysis shows that the separation of the variables \Eref{sepD}
due to Pokrovsky\cite{Pokrovskii:62} and consequent factorizable kernel
\Eref{sepV} are tools to apply the weak-coupling BCS approximation to
anisotropic-gap superconductors. The factorizable kernel gives a simple
solution to the gap equation, the nontrivial detail being that this
separability can be derived by the BCS gap equation. The factorizable kernel
has also been discussed by Markowitz and Kadanoff\cite{Markowitz:65} and
employed, e.g., by Clem\cite{Clem:66} to investigate the effect of gap
anisotropy in pure and superconductors with nonmagnetic impurities.
Factorizable kernels are now used in many works on exotic superconductors.
However, in none of them is mentioned that the separability of the
superconducting order parameter is an immanent property of the BCS
theory.\cite{Pokrovskii:62} The accuracy of the separable approximation is
higher if the other eigenvalues of the pairing kernel are much smaller than
the maximal one. This is likely to be the situation for the $s$-$d$ model for
layered cuprates, \cite{Mishonov:03a} where the $s$-$d$ pairing amplitude
$J_{sd}$ is much bigger than the phonon attraction and the other interatomic
exchange integrals. In order for us to clarify this important approach to the
theory of superconductivity, we have given here a rather methodical derivation
of the Pokrovsky theory.

%---------------------------------------------------------

\section{C++ code for computing the $\bm{F(x)}$ function
  E\lowercase{q}.~(\ref{Ffunc})}
\label{CppAppendix}

\renewcommand{\baselinestretch}{0.8}
\footnotesize

\begin{verbatim}
/*
 *  Calculates the F(x) function. By default this function
 *  uses the epsilon algorithm (the limes() function), but if
 *  you need faster calculation with smaller accuracy, you
 *  can set the optimized argument to false. Uses sum for
 *  x < 0.32, hankel sum for x >= 0.32, the
 *  F(x) ~= sqrt(2PI/x).exp(-x) approximation for x > 13
 *  and F(x) ~= log(PI / (GAMA * x)) for x < 0.001
 */
double F(double x, bool optimized = true) {
    if(x == 0) // return maximum possible double value.
    {
        return 1.0E308;
    }

    if(x < 0) // F(x) is an even function.
    {
        x = -x;
    }

    // For x < 0.001 we use the approximation
    // F(x) ~= log(PI / (GAMA * x)).
    if(x < 0.001)
    {
        return log(PI / (GAMA * x));
    }

    // For x > 13 we use the approximation
    // F(x) ~= sqrt(2PI/x).exp(-x).
    if(x > 13.0)
    {
        return sqrt(2.0*PI/x) * exp(-x);
    }

    double F_limes(double);
    double F_fast(double);

    return optimized ? F_limes(x) : F_fast(x);
}

/*
 *  Calculates the F(x) function without resorting to
 *  the epsilon algorithm.
 */ 
double F_fast(double x) {
    double sum=0.0, oldSum, k=1.0, eps, PI2=PI*PI;

    // For 0.001 <= x < 0.32 we use the sum
    // 0.5*F(x)  = 0.5 * log(PI / (GAMA * x)) +
    //                1 - PI/sqrt(x^2 + PI^2) +
    //           1/3 - PI/sqrt(x^2 + (3PI)^2) +
    //       ................................ +
    // 1/(2l+1) - PI/sqrt(x^2 + ((2l+1)PI)^2) +
    //       ................................ +
    if(x < 0.32)
    {
        eps = 2.0E-8 / (x * sqrt(x));
        sum = 0.5 * log(PI / (GAMA * x));
        do
        {
            oldSum = sum;
            sum += 1.0 / k - PI / (sqrt(x*x+k*k*PI2));
            k += 2.0;
        }
        while(fabs(sum - oldSum) >= eps * fabs(sum));

        return 2.0 * sum;
    }

    // For 0.32 <= x <= 13 we use the sum
    // 0.5*F(x) = K0(x) - K0(2x) + K0(3x) - ...
    // + (-1)^(n+1)*K0(nx) + ...
    eps = 2.3E-6 * pow(x, 0.8) * exp(1.5 * (x - 0.3));
    do
    {
        oldSum = sum;
        sum += bessk0(k * x) - bessk0((k + 1.0) * x);
        k += 2.0;
    }
    while(fabs(sum - oldSum) >= eps * fabs(sum));

    return 2.0 * sum;
}

/*
 *  Calculates the F(x) function using the epsilon algorithm.
 */
double F_limes(double x) {
    double err;
    int iPade, kPade;
    // Number of the partial sums.
    int N = 11;
    // Array, storing the partial sums.
    double Sn[12];
    // Array, storing the limes() function result.
    double An[12];
    double k = 1.0;
    // The lower bound should start at 0!
    Sn[0] = 0.0;
    // For 0.001 <= x < 0.32 we use the sum
    // 0.5*F(x)  = 0.5 * log(PI / (GAMA * x)) +
    //                1 - PI/sqrt(x^2 + PI^2) +
    //           1/3 - PI/sqrt(x^2 + (3PI)^2) +
    //       ................................ +
    // 1/(2l+1) - PI/sqrt(x^2 + ((2l+1)PI)^2) +
    //       ................................ +
    if(x < 0.32)
    {
        double PI2 = PI * PI;
        for(int i=1; i<=N; i++)
        {
            Sn[i] = Sn[i-1] +
                1.0/k-PI/(sqrt(x*x + k*k*PI2));
            k += 2.0;
        }

        return log(PI / (GAMA * x)) +
            2.0*limes(Sn, An, N, err, iPade, kPade);
    }
    // For 0.32 <= x <= 13 we use the sum
    // 0.5*F(x) = K0(x) - K0(2x) + K0(3x) - ...
    // + (-1)^(n+1)*K0(nx) + ...
    for(int i=1; i<=N; i++)
    {
        Sn[i] = Sn[i-1] +
            bessk0(k*x) - bessk0((k + 1.0)*x);
        k += 2.0;
    }

    return 2.0*limes(Sn, An, N, err, iPade, kPade);
}

/*
 *  Finds the limit of a series in the case where only 
 *  the first N+1 terms are known. Method: The subroutine 
 *  operates by applying the epsilon-algorithm to the 
 *  sequence of partial sums of a series supplied on input.
 *  For further details, please see:
 *  T. Mishonov and E. Penev
 *  Int. J. Mod. Phys. B 14, 3831 (2000). 
 */
double limes(double* S, double* A, int N,
             double& err, int& i_pade, int& k_pade)
{
    int i, k = 1;
    double rLimes = S[N], A_max = 0.0;
    err = fabs(S[N] - S[N-1]);
    i_pade = N;
    k_pade = 0;
    for(i=0; i<=N; i++)
    {
        A[i] = 0.0;
    }

    while(N - 2*k + 1 >= 0)
    {
        for(i=0; i<=N - 2*k + 1; i++)
        {
            A[i] = (S[i+1] != S[i]) ?
                A[i+1] + 1.0 / (S[i+1] - S[i] : A[i+1];
        }

        if(N - 2*k < 0)
        {
            break;
        }

        for(i=0; i<=N - 2*k; i++)
        {
            S[i] = (A[i+1] != A[i]) ?
                S[i+1] + 1.0 / (A[i+1] - A[i]) : S[i+1];

            if(fabs(A[i]) > A_max)
            {
                A_max = fabs(A[i]);
                rLimes = S[i];
                k_pade = k;
                i_pade = i + k_pade;
                err = 1.0/A_max;
                if(S[i+1] == S[i])
                {
                    return rLimes;
                }
            }
        }

        k++;
    }

    return rLimes;
}
\end{verbatim}

\end{appendix}

%\bibliographystyle{apsrev}
%\bibliography{heat}

\end{document}